\documentclass[12pt]{article}

   \usepackage{amsmath}
\usepackage{graphicx,psfrag,epsf}
\usepackage{enumerate}
\usepackage{natbib}
\usepackage{url} 
\usepackage{comment}
\newcommand{\blind}{1}
\usepackage{xr}
\usepackage{amsmath,amssymb,amsfonts,bbm}
\usepackage{multirow,verbatim}
\usepackage{amsthm}
\def\SMB{\mbox{SMB}}
\def\HML{\mbox{HML}}

\def\RMSE{\mbox{RMSE}}

\renewcommand{\hat}{\widehat}

\newcommand{\bfm}[1]{\ensuremath{\mathbf{#1}}}

   \def\bA{\bfm A}  
\def\bb{\bfm b}   \def\bB{\bfm B}  
   \def\bC{\bfm C}  
   \def\bD{\bfm D}  
\def\be{\bfm e}     
\def\bff{\bfm f}    
\def\bg{\bfm g}   \def\bG{\bfm G}  
\def\bh{\bfm h}   \def\bH{\bfm H}  
   \def\bI{\bfm I}

   \def\bL{\bfm L}  
   \def\bM{\bfm M}

   \def\bP{\bfm P}  
     
   \def\bR{\bfm R}

\def\bu{\bfm u}     
\def\bv{\bfm v}   \def\bV{\bfm V}  
\def\bw{\bfm w}   \def\bW{\bfm W}  
\def\bx{\bfm x}     
\def\by{\bfm y}   \def\bY{\bfm Y}  
\def\bz{\bfm z}

\newcommand{\bfsym}[1]{\ensuremath{\boldsymbol{#1}}}

 \def\balpha{\bfsym \alpha}
 \def\bbeta{\bfsym \beta}
 \def\bgamma{\bfsym \gamma}             
            \def\bDelta {\bfsym {\Delta}}
               
 \def\bmu{\bfsym {\mu}}                 
 \def\bnu{\bfsym {\nu}}
            
           \def\bepsilon{\bfsym \varepsilon}
              \def\bSigma{\bfsym \Sigma}
 \def\blambda {\bfsym {\lambda}}        \def\bLambda {\bfsym {\Lambda}}

 \def\bxi{\bfsym {\xi}}
 
  \def\bXi{\bfsym {\Xi}}

\DeclareMathOperator{\cov}{cov}

\DeclareMathOperator{\diag}{diag}

\DeclareMathOperator{\sgn}{sgn}

\DeclareMathOperator{\var}{var}

\DeclareMathOperator{\tr}{tr}
\numberwithin{equation}{section}
\theoremstyle{plain}
\newtheorem{thm}{Theorem}[section]
\newtheorem{defn}{Definition}[section]

\newtheorem{prop}{Proposition}[section]
\newtheorem{ass}{Assumption}[section]

\theoremstyle{definition}

\newtheorem{rem}{Remark}[section]

\begin{document}

\def\spacingset#1{\renewcommand{\baselinestretch}%
{#1}\small\normalsize} \spacingset{1}

\if1\blind
{
  \title{\bf Augmented Factor Models with Applications to Validating Market Risk Factors and Forecasting Bond Risk Premia
  \thanks{Fan gratefully acknowledges the support of NSF grant DMS-1712591.}}
  \author{Jianqing Fan$^+$, Yuan Ke$^\dag$ and Yuan Liao$^\ddag$
  \hspace{.2cm}\\
   $^+$Princeton University, $^\dag$University of Georgia, $^\ddag$ Rutgers University
        }
  \maketitle
} \fi

\if0\blind
{
  \bigskip
  \bigskip
  \bigskip
  \begin{center}
    {\LARGE\bf Augmented Factor Models with Predictions and Inferences }
\end{center}
  \medskip
} \fi

\bigskip
\begin{abstract}
We study   factor models augmented by observed covariates that have explanatory powers on the unknown factors.   In financial factor models, the unknown factors can be reasonably well explained by a few observable proxies, such as the Fama-French factors. In diffusion index forecasts, identified factors are strongly related to several directly measurable economic variables such as consumption-wealth variable, financial ratios, and term spread.
With those covariates, both the factors and loadings are identifiable up to a rotation matrix even only with a  finite dimension.    To incorporate the explanatory power of these   covariates, we propose a smoothed principal component analysis (PCA):  (i) regress the data onto the observed covariates, and (ii) take the principal components of the fitted data to estimate the loadings and factors.    This allows us to accurately estimate the percentage of both explained and unexplained components in factors and thus to assess the explanatory power of covariates.
  We show that both the estimated factors and loadings can be estimated with improved rates of convergence compared to the benchmark method. The degree of improvement depends on the strength of the signals, representing the explanatory power of the covariates on the factors. The proposed estimator is robust to possibly heavy-tailed distributions.   We apply the model to forecast US bond risk premia, and find that the observed macroeconomic characteristics  contain strong explanatory powers of the factors. The gain of forecast is more substantial when the characteristics are incorporated to estimate the common factors  than directly used for forecasts.
\end{abstract}

\noindent%


\spacingset{1.45} 

\section{Introduction}
\label{sec:introduction}

In this paper, we study the identification and estimations of factor models augmented by a set of additional covariates that are common to all individuals.
Consider the following factor model:
\begin{equation} \label{e2.1}
\by_t=\bLambda \bff_t+\bu_t,\quad t = 1, \cdots, T.
\end{equation}
Here  $\by_t=(y_{1t},..., y_{Nt})'$ is the multivariate outcome for the $t^{th}$ observation in the sample; $\bff_t$ is the $K$-dimensional vector of latent factors; $\bLambda=(\blambda_1,....,\blambda_N)'$ is an $N\times K$ matrix of nonrandom factor loadings;  $\bu_t=(u_{1t},...,u_{Nt})'$ denotes the  vector of idiosyncratic errors. In addition to  $\{\by_t\}_{t=1}^T$,  we also  observe  variables, denoted by $\bx_t$,  that have some explanatory power on the unknown factors and hence impact  on observed vector $\by_t$. We  model $\bff_t$ by using the model
\begin{equation}\label{e1.2}
\bff_t=\bg(\bx_t)+\bgamma_t,
\end{equation}
for some (nonparametric) function $\bg = E(\bff_t|\bx_t)$. Here $\bg(\bx_t)$ is interpreted as the component of the factors that can be explained by the covariates, and  $\bgamma_t$ is the components that cannot be explained by the covariates.
We aim to provide an improved estimation procedure when the factors can be partially explained by several observed  variables $\bx_t$.
In addition, by accurately estimating $\bgamma_t$, we can  estimate the percentage of both explained and unexplained components in the factors, which describes the proxy/explanatory power of covariates.


Note that  model (\ref{e2.1}) implies:
 \begin{equation} \label{e1.3}
 \cov(\by_t)=\bLambda \cov(\bff_t)\bLambda' +\cov(\bu_t),
 \end{equation}
 where $\cov(\by_t)$ and $\cov(\bu_t)$ respectively denote the $N\times N$ variance-covariance matrices of $\by_t$ and $\bu_t$; $\cov(\bff_t)$ denotes the $K\times K$  variance-covariance matrix of $\bff_t$. Under usual factor models without covariates,  $\frac{1}{\sqrt{N}}\bLambda$ is  identified \textit{asymptotically} as the first $K$ eigenvectors of $\cov(\by_t)$ as $N\to\infty$ and can be estimated using the first $K$ eigenvectors of the sample covariance matrix of $\by_t$ (e.g,, \cite{SW02, bai03}).

With additional covariates, on the other hand, \textit{exact} identification can be achieved  through covariance of the ``smoothed data''.  By \eqref{e2.1}, assuming exogeneity of $\bx_t$, we have
$
    E(\by_t|\bx_t) = \bLambda E(\bff_t | \bx_t)
$
so that it becomes a ``noiseless" factor model with smoothed data 
$ E(\by_t|\bx_t)$ as input and $E(\bff_t | \bx_t)$ as latent factors.  The factor loadings and latent factors can be extracted from
\begin{equation}\label{a4}
\bSigma_{y|x}=E\{E(\by_t|\bx_t) E(\by_t|\bx_t)'\}.
\end{equation}
It is easy to see from the model that
\begin{equation} \label{a5}
\bSigma_{y|x} =\bLambda\bSigma_{f|x} \bLambda',
\end{equation}
where  $\bSigma_{f|x} = E\{E(\bff_t|\bx_t)E(\bff_t|\bx_t)'\}$  is   a $K\times K$ low-dimensional positive definite matrix. This decomposition is to be compared with (\ref{e1.3}), where the noise covariance $\cov(\bu_t)$ removed.   Therefore, as long as
$\bSigma_{f|x}$ is of full rank, $\bLambda$ falls in the eigenspace generated by $\bSigma_{y|x}$.  In other words, $\bLambda$ is identifiable up to an orthogonal transformation. Because of such exact identification, we allow $N$ to be  finite as a special case.  
The number  of factors is assumed to be known throughout the paper. In practice, $K$ can be consistently estimated by many methods such as AIC, BIC-based criteria, or eigenvalue-ratio methods studied in \cite{LamYao,AH}.

The above discussion prompts  us  the following new method to estimate the factor loadings $\bLambda$ that incorporates the explanatory power of $\bx_t$: (See Section 3 for details of estimators)

(i)  (robustly) regress $\{\by_t\}$ on  $\{\bx_t\}$ and obtain fitted value $\{\widehat\by_t\}$;

(ii) conduct the principal components analysis (PCA) on the fitted data $(\widehat\by_1,...,\widehat\by_T)$ to estimate the factor  loadings.
\\
 We employ  a   regression based on   \cite{huber1964robust}'s robust M-estimation in step (i). The procedure involves a diverging truncation parameter,  called adaptive Huber loss, to reduce the bias when the error distribution is asymmetric \citep{fan2017estimation}.
 This allows our procedure to be applicable  to data with heavy tails.\footnote{In this paper, by ``heavy-tail" we mean tail distributions of $(\bu_t, \by_t)$ that are heavier than the usual requirements on the high-dimensional factor model    (which are either exponentially-tailed or have eighth or higher moments). But we do not allow large outliers on the covariates. } 


There are two important quantities that determine the rates of convergence for the estimators:
the ``signal"  $\bSigma_{f|x}= E\{E(\bff_t|\bx_t)E(\bff_t|\bx_t)'\}$ and the ``noise" $\cov(\bgamma_t)$. The rates of convergence are presented using these two quantities. Their relative strengths determine the rates of convergence of the estimated factors and loadings.

Under model (\ref{e1.2}), we can test $\bgamma_t=0$ almost surely in the entire sampling period,  under which the observed $\bx_t$ fully explain the true factors.  This is the same as testing
$$
H_0: \cov(\bgamma_t)=0.
$$
While it is well known that the commonly used Fama-French factors have explanatory power for most of the variations of stock returns, it is questionable whether they fully explain the true (yet unknown) factors. These observed proxies are nevertheless used as the factors empirically, and the remaining components ($\bgamma_t$ and $\bu_t$) have all been mistakenly regarded as the idiosyncratic components.
The proposed test provides a diagnostic tool for the specification of common factors in empirical studies, and is different from the ``efficiency test" in the financial econometric literature (e.g., \cite{GRS, PY, gungor2013testing}). While the efficiency test aims to test the asset pricing model through whether the alphas are zero for the specified factors,  a rejection could be due to either mispecified factors or the existence of outperforming (underperforming) assets. 
In contrast, here we directly test whether the factor proxies are correctly specified. We test the specification of Fama French factors for the returns of S\&P 500 constituents using rolling windows. We find that the null hypothesis is more often to be rejected using the daily data compared to the monthly data, due to a larger volatility of the unexplained factor components. The estimated overall volatility of factors varies over time and drops significantly during the acceptance period.


\subsection{Further Literature}
In empirical applications, researchers frequently encounter additional observable covariates that help explain the latent factors.  In genomic studies, in  the study of  breast cancer data such as  the Cancer Genome Atlas (TCGA) project \citep{cancer2012comprehensive},  there are additional information of cancer subtype for each sample. These  cancer subtypes can be regarded as a
partial driver of the  factors for gene expression data.  In financial time series forecasts,  researchers often collect additional variables that characterize   financial markets. The Fama-French factors are well-known to be related to the factors that drive financial returns \citep{FF}.  

Most  existing works   simply treat $\bx_t$ as a set of  additional regressors in (\ref{e2.1}). This approach does not take advantage of the difference of observed variables (e.g. aggregated versus disaggregated macroeconomic variables; gene expressions versus clinical information)  and the explanatory power of the covariates on the common factors, and hence does not lead to  improved rates of convergence even if the signal is strong. 
The most related work is \cite{li2016supervised}, who specified $\bff_t$ as a linear function of $\bx_t$.  Also, \cite{huang2010combine}  proposed to use the estimated $\bg(\bx_t)$ to forecast.  Moreover, our expansion is also connected to the literature on   asymptotic Bahadur-type representations for robust M-estimators, see, for example, \cite{portnoy1985asymptotic},    \cite{mammen1989asymptotics},  among others.


The ``asymptotic identification" was    described perhaps first by  \cite{CR}. In addition, there has been a large literature  on both the static and dynamic factor models, and we refer to    \cite{Lawley,   Forni05, SW02, BN02, bai03, doz, onatski, poet},   among many others.

 The rest of the paper is organized as follows. Section 2  establishes  the new identification of factor models. Section \ref{sec:est}  formally defines our estimators and discusses   possible alternatives. Section \ref{sec3} presents the  rates of convergence.  Section \ref{sec:Test} discusses the problem of testing the explanatory power. Section \ref{sec:Emp1} applies the model to forecasting the excess return of US government bonds.  We present the extensive simulation studies in Section \ref{sec:Sim}   Finally Section \ref{sec:con} concludes.  The  supplement also contains   all the technical proofs.


 Throughout the paper, we use $\lambda_{\min}(\bA)$ and $\lambda_{\max}(\bA)$ to denote the minimum and maximum eigenvalues of a matrix $\bA$. We  define $\|\bA\|_F=\tr^{1/2}(\bA'\bA)$, $\|\bA\|=\lambda_{\max}^{1/2}(\bA'\bA)$, $\|\bA\|_1 =\max_{j} \sum_{i} |a_{ij}|$ and $\|\bA\|_{\max}=\max_{i,j}|a_{ij}|$. For two sequences, we write $a_T\gg b_T$ or $b_T\ll a_T$ if $b_T=o(a_T)$ and $a_T\asymp b_T$ if  $a_T=O(b_T)$ and $b_T=O(a_T).$

\section{Identification of the covariate-based factor models}

 \subsection{Identification}
Suppose that there is a fixed $d$-dimensional observable vector $\bx_t$ that is: (i)     associated with the latent factors $\bff_t$, and (ii) mean-independent of  the idiosyncratic term.    Taking the conditional mean  on both sides of  (\ref{e2.1}), we have
\begin{equation}\label{e2.2}
E(\by_t|\bx_t)= \bLambda E(\bff_t|\bx_t),
\end{equation}
This implies
\begin{equation}\label{e2.3}
\bSigma_{y|x}=\bLambda \bSigma_{f|x}\bLambda',
\end{equation}
where
\begin{eqnarray*}
\bSigma_{y|x} :=E\{  E(\by_t|\bx_t) E(\by_t|\bx_t)' \},\qquad
\bSigma_{f|x} :=E\{E(\bff_t|\bx_t)E(\bff_t|\bx_t)'\}.
\end{eqnarray*}
Note that  $ E(\by_t|\bx_t) $ is identified by the data generating process with observables $\{(\by_t, \bx_t)\}_{t\leq T}$, but   $\bSigma_{f|x}$ is not  because $\bff_t$ is not observable.
Since  $N>K$,   (\ref{e2.3})  implies that $\bSigma_{y|x}$  is a low-rank matrix, whose rank is at most $K.$
Furthermore, we assume $ \bSigma_{f|x}$ is also full rank, so $\bSigma_{y|x}$ has exactly $K$ nonzero eigenvalues.

To see how the equality (\ref{e2.3}) helps  achieve the identification of $\bLambda$ and $\bg(\bx_t)$,
for the moment, suppose the following normalization   holds:
\begin{equation}\label{e2.3add}
\frac{1}{N}\bLambda'\bLambda=\bI_K, \quad
  \bSigma_{f|x}  \text{ is a diagonal matrix.}
  \end{equation}
   Then  right multiplying (\ref{e2.3}) by $\bLambda/N$, by the normalization condition,
$$
\frac{1}{N}\bSigma_{y|x}\bLambda =\bLambda \bSigma_{f|x}.
$$
We see that  the ($K$) columns of $\frac{1}{\sqrt{N}}\bLambda$   are  the eigenvectors of $\bSigma_{y|x}$, corresponding to its $K$ nonzero eigenvalues, which also equal to the diagonal entries   of $ \bSigma_{f|x} $.
Furthermore, left multiplying $\bLambda'/N$ on both sides of (\ref{e2.2}), one can see that  even if $\bff_t$ is not observable, $E(\bff_t|\bx_t)$ is also identified as:
$$
\bg(\bx_t):=E(\bff_t|\bx_t)=\frac{1}{N}\bLambda' E(\by_t|\bx_t).
$$

The normalization (\ref{e2.3add}) above is useful to facilitate the above arguments. In this paper, they are not imposed. Then the   same argument shows  that $\bLambda$ and $\bg(\bx_t)$ can be identified up to  a rotation matrix transformation.  

Let \begin{eqnarray*}
\bSigma_{\Lambda, N}&:=&\bLambda'\bLambda/N, \qquad \qquad
 \chi_N:=\lambda_{\min}(E\{E(\bff_t|\bx_t)E(\bff_t|\bx_t)'\}).
\end{eqnarray*}
\begin{ass}\label{a1} Suppose $\{\bff_t,\bx_t,\bu_t\}_{t\leq T}$ are identically distributed.  Assume:

(i)  Rank condition:  $\chi_N>0$.

(ii) There are positive constants $\underbar{c}_\Lambda, \bar{c}_\Lambda>0$, so that all the eigenvalues of the $K\times K$ matrix $\bSigma_{\Lambda,N}$ are confined in $[\underbar{c}_\Lambda, \bar{c}_\Lambda]$, regardless of whether $N\to\infty$ or not.


\end{ass}

Condition (i) is the key condition on the  explanatory power of $\bx_t$ on factors, where $\chi_N$ represents the ``signal strength" of the model. We  postpone   the discussion of this condition  after Theorem \ref{th0}.
 Condition (ii) in Assumption  \ref{a1} can be weakened  to allow the eigenvalues of $\bSigma_{\Lambda, N}$ to slowly decay to zero. While doing  so allows some of the factors to be \textit{weak},   it does not provide any new statistical insights,  but would bring unnecessary complications to our  results and conditions.  Therefore, we maintain  the strong version as  condition (ii).


 Generally, we have  the following theorem for identifying $(\bLambda, \bg(\bx_t))$ (up to a rotation transformation).

 \begin{thm}\label{th0}
Suppose $E(\bu_t|\bx_t)=0$,
Assumption \ref{a1} holds and $N>K$. Then there is an invertible $K\times K$ matrix  $\bH$ so that:

(i)  The columns of $\bLambda \bH$ are the eigenvectors of $\bSigma_{y|x}$  corresponding to the nonzero distinct eigenvalues.

(ii) Given $\bLambda\bH$, $\bg(\bx_t):=E(\bff_t|\bx_t) $ satisfies:
$$
\bH^{-1}\bg(\bx_t)=[(\bLambda\bH)'\bLambda\bH ]^{-1}\bLambda\bH'E(\by_t|\bx_t).
$$

(iii) Let $\lambda_K(\bSigma_{y|x})$ denote the $K$th largest eigenvalue of $\bSigma_{y|x}$, we have
$$
\lambda_K(\bSigma_{y|x})\geq   N\chi_N\underbar{c}_\Lambda.
$$
where $\chi_N$ and $\underbar{c}_\Lambda$ are defined in Assumption \ref{a1}.
  In addition, under the normalization conditions that $ E\{E(\bff_t|\bx_t)E(\bff_t|\bx_t)'\} $  is a diagonal matrix and that $ \bSigma_{\Lambda,N}=\bI_K$, we have
$
\bH=\bI_K.
$

  \end{thm}




  \subsection{Discussions of Condition (i) of Assumption \ref{a1}}
 In the model $$
  \bff_t=\bg(\bx_t)+\bgamma_t,\quad \bg(\bx_t)=E(\bff_t|\bx_t),
  $$
   $ \chi_N =\lambda_{\min}(\bSigma_{f|x})$   represents the
  ``signal" of the covariate model.  We require $\chi_N>0$ so that the rank of $\bSigma_{y|x}$ is $K$.
Only if this condition holds are we  able to     identify all the $K$ factor loadings using the   eigenvectors corresponding to the  nonzero eigenvalues. 
 From the estimation point of view, we are using the PCAs of the estimated $\bSigma_{y|x}$, and  can only consistently estimate its rank$(\bSigma_{y|x})$-number of leading eigenvectors. So this condition is also essential to achieve the consistent estimation of the factor loadings.

Note that requiring $\bSigma_{f|x}$ be of full rank might be  restrictive in some cases. For instance, consider the linear  case:
$
E(\bff_t|\bx_t) =\bbeta\bx_t$ for a $K\times d$ coefficient matrix $\bbeta$, also suppose $E\bx_t\bx_t'$ is of full rank. Then
$\bSigma_{f|x}=\bbeta E\bx_t\bx_t'
\bbeta'$, and  is  full-rank only if $d\geq K$. Thus   we implicitly require,  for linear models,   the number of covariates should be at least as many as the number of latent factors.   
Note that  if $E(\bff_t|\bx_t)$ is highly nonlinear, it is still possible  to satisfy the full rank condition  even if $d<K$, and we illustrate this in the simulation section. \footnote{Suppose  $E(\bff_t|\bx_t)$ is nonlinear and  can be well approximated by a series of orthogonal basis functions $\Phi(\bx_t)=(\phi_1(\bx_t),...,\phi_J(\bx_t))'$, where $E\phi_i(\bx_t)\phi_j(\bx_t)=1\{i=j\}$, then for some $K\times J$ coefficient $\balpha$, we have $E(\bff_t|\bx_t)\approx \balpha' \Phi(\bx_t)$ so
 $E\{E(\bff_t|\bx_t)E(\bff_t|\bx_t)'\}\approx \balpha\balpha'$.  For  nonlinear functions, it is not stringent to require  $\balpha\balpha'$  be full rank since $K<J$ as $J\to\infty$.
}

\section{Definition of the estimators}
 \label{sec:est}

The above identification strategy motivates us to estimate $\bLambda$ and $\bg(\bx_t)$ respectively by $\widehat{\bLambda}$ and $\widehat \bg(\bx_t)$ as follows.
Let  $\widehat{\bSigma}$ and  $\widehat E(\by_t|\bx_t)$ be some  estimator of $\bSigma_{y|x}$ and $E(\by_t|\bx_t)$, whose definitions will be clear below.   Then the columns of  $\frac{1}{\sqrt{N}}\widehat{\bLambda}$ are defined as the eigenvectors corresponding to the first $K$ eigenvalues of  $\widehat{\bSigma}$, and
$$\widehat\bg(\bx_t):=\frac{1}{N}\widehat{\bLambda}' \widehat E(\by_t|\bx_t).
$$
Recall that $\bff_t=\bg(\bx_t)+\bgamma_t$.
We estimate $\bff_t$   using least squares:
$$
\widehat\bff_t:=(\widehat{\bLambda}'\widehat{\bLambda})^{-1}\widehat{\bLambda}'\by_t=\frac{1}{N}\widehat{\bLambda}' \by_t.
$$
Finally, we estimate $\bgamma_t$ by:
$
\widehat{\bgamma}_t=\widehat\bff_t-\widehat\bg(\bx_t)=\frac{1}{N}\widehat{\bLambda}'(\by_t-\widehat E(\by_t|\bx_t)).
$
Estimating $\bg(\bx_t)$ and $\bgamma_t$ separately allows us to estimate and distinguish the percentage of explained and unexplained components in factors, as well as to quantify the explanatory power of covariates.

Below we introduce the estimators  $\widehat{\bSigma}$ and  $\widehat E(\by_t|\bx_t)$  to be used in this paper.

\subsection{Robust estimation for $\widehat{\bSigma}$}\label{Sieve_LS}

   Recall that $\bSigma_{y|x}=E\{  E(\by_t|\bx_t) E(\by_t|\bx_t)' \}$, and let us    first construct an estimator for $ E(\by_t|\bx_t)$   as follows.  While many standard nonparametric regressions would work, here we choose an estimator that is robust to the tail-distributions of $\by_t-E(\by_t|\bx_t)$.

Let $\Phi(\bx_t)=(\phi_1(\bx_t),...,\phi_J(\bx_t))'$ be a $J\times 1$ dimensional
 vector of sieve basis.
 Suppose $E(\by_t|\bx_t)$ can be approximated by a sieve representation:
$
E(\by_t|\bx_t)\approx \bB \Phi(\bx_t),
$
where $\bB=(\bb_1,...,\bb_N)'$ is an $N\times J$ matrix of sieve coefficients.
To adapt to different heaviness of the tails of idiosyncratic components, we  use the Huber loss function (\cite{huber1964robust}) to estimate the sieve coefficients. Define
$$
\rho(z)=\begin{cases} z^2, & |z|<1\\
2|z|-1, & |z|\geq 1.
\end{cases}
$$
For some deterministic sequence $\alpha_T\to\infty$ (adaptive Huber loss),  we estimate the sieve coefficients $\bB$ by the following convex optimization:
$$
\widehat{\bb}_i=\arg\min_{b\in\mathbb{R}^J}   \frac{1}{T}\sum_{t=1}^T \rho\left( \frac{y_{it}-\Phi(\bx_t)'\bb}{\alpha_T}\right),\quad \widehat{\bb}=(\widehat{\bb}_1,...,\widehat{\bb}_N)'.
$$
We then estimate $\bSigma_{y|x}$ by
$$
 \widehat{\bSigma}=\frac{1}{T}\sum_{t=1}^T\widehat E(\by_t|\bx_t) \widehat E(\by_t|\bx_t)',\quad \text{where } \widehat E(\by_t|\bx_t)=\widehat{\bb}\Phi(\bx_t).
 $$

An alternative method to the robust estimation of $\bSigma_{y|x}$ is based on the sieve-least squares,  corresponding to the case where $\alpha_T = \infty$. Let $\bY=(\by_1,...,\by_T), $ which is   $(N\times T)$, and
 $$
 \bP=\Phi'(\Phi\Phi')^{-1}\Phi,  (T\times T),\quad \Phi=(\Phi(\bx_1),...,\Phi(\bx_T)),   (J\times T).
 $$
Then, the  sieve least-squares estimator   for $\bSigma_{y|x}$ is
$
\widetilde\bSigma= \frac{1}{T}   \bY\bP\bY'$. 
 While this estimator is attractive due to its closed form,   it is not as good as $\widehat\bSigma$ when the  distribution of $\bu_t$ has heavier tails.  As expected, our numerical studies in Section \ref{sec:Sim}  demonstrate that it performs well in light-tailed scenarios, but is less  robust to heavy-tailed distributions.   Our theories are presented for $\widehat\bSigma$, but most of the theoretical findings should carry over to $\widetilde\bSigma$. 



\subsection{Choosing $\alpha_T$ and $J$}\label{sec_tunning}

The selection of the sieve dimension $J$ has been widely studied in the  literature, e.g.,  \cite{Li_87, Andrews_91, np_AIC_98}, among others.  Another tunning parameter is  $\alpha_T$, which     diverges in order to reduce the biases of estimating the conditional mean  when the distribution of $\by_t-E(\by_t|\bx_t)$ is asymmetric.    Throughout the paper, we shall set
\begin{equation}\label{alpha_T}
\alpha_T=C_\alpha\sqrt{\frac{T}{\log(NJ)}}
\end{equation}
for some constant $C_\alpha>0$, and choose $(J, C_\alpha)$ simultaneously using the multi-fold cross-validation\footnote{One can also allow $\alpha_T$ to depend on $\var(y_{it}|\bx_t)$ to allow for different scales across individuals. We describe this choice in the simulation section. In addition, the cross-validation can be based on either in-sample fit for $E(y_{it}|\bx_t)$ or out-of-sample forecast, depending on the specific applications. In time series forecasts, one may also consider the time series cross validation \citep[e.g.][]{TSCV} where the training and testing sets are defined through a moving window forecast. }.    The specified rate in (\ref{alpha_T}) is due to a theoretical consideration, which  leads to the ``least biased robust estimation", as we now explain.
The Huber-estimator is biased for estimating the mean coefficient in $E(y_{it}|\bx_t)$, whose population counterpart   is  $$
\bb_{i,\alpha}:=\arg\min_{\bb\in\mathbb{R}^J} E \rho\left( \frac{y_{it}-\Phi(\bx_t)'\bb}{\alpha_T}\right),
$$
As $\alpha_T$ increases,   it approaches the limit $\bb_i:=\arg\min_{\bb\in\mathbb{R}^J} E[y_{it}-\bb'\Phi(\bx_t)]^2$ with the speed
$$
\max_{i\leq N}\|\bb_{i,\alpha}-\bb_i\|=O(\alpha_T^{-(\zeta_2+1)+\epsilon})
$$
 for an arbitrarily small $\epsilon>0$, where $\zeta_2$ is defined in Assumption \ref{a31}. Hence the bias decreases as $\alpha_T$ grows. On the other hand, our theory requires the uniform convergence (in $i=1,...,N$) of (for $e_{it}=y_{it}-E(y_{it}|\bx_t)$)
\begin{equation}\label{eq3.1}
\max_{i\leq N}\|\frac{1}{T}\sum_{t=1}^T\dot{\rho}(\alpha_T^{-1}e_{it})\Phi(\bx_t)\|,
\end{equation}
where $\dot\rho(\cdot) $ denotes the derivative of $\rho(\cdot)$.     It turns out that  $\alpha_T$ cannot grow  faster than $O(\sqrt{\frac{T}{\log(NJ)}})$ in order to guard for robustness and  to have a sharp uniform convergence for (\ref{eq3.1}).  Hence  the choice (\ref{alpha_T}) leads to the  asymptotically least-biased robust estimation.

\subsection{Alternative estimators}




   Plugging $\bff_t=\bg(\bw_t)+\bgamma_t$ into (\ref{e2.1}), we obtain
\begin{equation} \label{e2.5add}
 \by_t=\bh(\bx_t)+\bLambda\bgamma_t+\bu_t,\quad\text{where } \bh(\bx_t)=\bLambda \bg(\bx_t).
\end{equation} A closely related model is:
 \begin{equation}\label{e2.6}
 \by_t=\bh(\bx_t) +\bLambda\bff_t+\bu_t,
 \end{equation}
  for a nonparametric function $\bh(\cdot)$, or simply a linear form $\bh(\bx_t)=\bbeta\bx_t$.  Models (\ref{e2.5add}) and (\ref{e2.6}) were studied  in the literature  \citep{ALS, bai, MW11}, where parameters are often estimated using least squares. For instance, we can estimate  model (\ref{e2.5add}) by
   \begin{equation}\label{e2.7}
 \min_{ \bh, \bLambda, \bgamma_t}\frac{1}{T}\sum_{t=1}^T\|\by_t-\bh(\bx_t)-\bLambda\bgamma_t\|^2.
 \end{equation}
 But this approach is not appropriate in the current context when $\bx_t$ almost fully explains $\bff_t$ for all $t=1,...,T$.  In this case, $\bgamma_t\approx 0$, and least squares (\ref{e2.7}) would be inconsistent. \footnote{The inconsistency is due to the fact that $a\bLambda\bgamma_t\approx \bLambda\bgamma_t$ for  \textit{any} scalar $a$ in the case  $\bgamma_t\approx0$. Thus $\bLambda$ is not identifiable in the least squares problem.}  In addition,  $\bLambda$  in (\ref{e2.6}) would be very close to zero because the effects of $\bff_t$ would be fully explained by $\bh(\bw_t)$. As a result, the factors in (\ref{e2.6}) cannot be consistently estimated \citep{onatski2012asymptotics} either. We conduct numerical comparisons with this method in the simulation section. In all simulated scenarios,  the interactive effect approach gives the worst estimation performance.

Another   simpler alternative  is to combine $(\bx_t, \by_t)$, and apply the classical methods on  this enlarged dataset. One potential drawback   is that the rates of convergence would not be improved, even if $\bx_t$ has strong explanatory power on the factors.    Another drawback, as mentioned before, is that $\bx_t$ and $\by_t$ can provide very different information (e.g. Fama-French factors versus returns of individual stocks).

 \section{Rates of Convergence}\label{sec3}

\subsection{Assumptions}





 Let
$
e_{it}:=y_{it}-E(y_{it}|\bx_t).
$
Suppose the conditional distribution of $e_{it}$ given $\bx_t=\bx$ is absolutely continuous for almost all $\bx$, with a conditional density $g_{e, i}(\cdot|\bx)$.
\begin{ass}[Tail distributions]\label{a31}
(i) There are $\zeta_1, \zeta_2>2$, $C>0$ and $M>0$, so that for all $x>M$,
\begin{equation}\label{e3.1}
\sup_{\bx}\max_{i\leq N} g_{e,i}(x|\bx)\leq Cx^{-\zeta_1},\quad \sup_{\bx}\max_{i\leq N} E(e_{it}^21\{|e_{it}|>x\}|\bx_t=\bx)\leq Cx^{-\zeta_2}.
\end{equation}
(ii)  $\Phi(\bx_t)$ is a  sub-Gaussian vector, that is, there is $L>0$, for any $\bnu\in\mathbb{R}^J$ so that $\|\bnu\|=1$,
$$
P(|\bnu'\Phi(\bx_t)|>x)\leq \exp(1-x^2/L),\quad\forall x\geq0.
$$
\end{ass}

\begin{ass}[Sieve approximations]\label{a3.3}
(i) For $k=1,...,K$, let $\bv_k=\arg\min_{\bv}E(f_{kt}-\bv'\Phi(\bx_t))^2$.  Then there is $\eta\geq 2$, as $J\to\infty$,
$$
\max_{k\leq K}\sup_{\bx}|E(f_{tk}|\bx_t=\bx)-\bv_k'\Phi(\bx)|=O(J^{-\eta}).
$$
(ii) There are $c_1, c_2>0$ so that
\begin{eqnarray*}
&&c_1\leq\lambda_{\min}(E\Phi(\bx_t)\Phi(\bx_t)')\leq\lambda_{\max}(E\Phi(\bx_t)\Phi(\bx_t)')\leq c_2.
\end{eqnarray*}

\end{ass}


Recall  $\bgamma_t=\bff_t-E(\bff_t|\bx_t)$. Let $\gamma_{kt}$ be its $k$ th component.

\begin{ass}[Weak dependences] \label{a32}

(i)  (serial independence)  $\{\bff_t, \bu_t, \bx_t\}_{t\leq T}$ is independent and identically distributed;

(ii) (weak cross-sectional dependence) For some $C>0$,
$$\sup_{\bx, \bff} \max_{i\leq N}\sum_{j=1}^N|E(u_{it}u_{jt}|\bx_t=\bx, \bff_t=\bff)| <C.$$
(iii) $ E(\bu_t|\bff_t, \bx_t)=0$,    $\max_{i\leq N}\|\blambda_i\|<C$, and $\cov(\bgamma_t|\bx_t)=\cov(\bgamma_t)$ almost surely, where $\cov(\bgamma_t|\bx_t)$ denotes the conditional covariance matrix of $\bgamma_t$ given $\bx_t$, assumed to exist.

\end{ass}

 Recall that
$$\bSigma_{f|x}:=E\{E(\bff_t|\bx_t)E(\bff_t|\bx_t)'\} ,\quad \chi_N:=\lambda_{\min}(     \bSigma_{f|x} ) .$$

\begin{ass}[Signal-noise]\label{asn}
(i) There is $C>0$,
$$
\frac{\lambda_{\max}( \bSigma_{f|x})}{\lambda_{\min}( \bSigma_{f|x} )}<C,
\quad
\frac{\lambda_{\max}( E\{\Phi(\bx_t)E(\bff_t|\bx_t)'E(\bff_t|\bx_t)\Phi(\bx_t)'\})}{\lambda_{\min}(\bSigma_{f|x}  )}<C.
$$
 (ii) There is $v>1$, so that $\max_{k\leq K}E[E(\gamma_{kt}^4|\bx_t)]^v<\infty$.\\
 (iii) We have $J^3\log ^2N=O(T)$ and $$ J^{2}/T+J^{-\eta}+\sqrt{(\log N)/T}\ll \chi_N. $$
\end{ass}




Assumption \ref{a31}  allows distributions with relatively heavy tails   on $y_{it}-E(y_{it}|\bx_t)$.  We   still   require sub-Gaussian tails for the sieve basis functions. 
Assumption \ref{a3.3} is regarding the accuracy of sieve approximations for nonparametric functions.   Assumption \ref{asn} strengthens  Assumption  \ref{a1}.   We respectively  regard $   \lambda_{\min}(\bSigma_{f|x})$  and $\cov(\bgamma_t)$  as the ``signal" and   ``noise" when  using $\bx_t$ to explain common factors.
 The explanatory power is measured by these two quantities. 

 Assumption \ref{a32} (i) requires  serial independence, and we admit that it can be  restrictive in applications.   Allowing for serial dependence is technically difficult due to the non-smooth Huber's loss. To obtan the      Bahadur representation   of the estimated eigenvectors,  we rely on the
symmetrization
 and  contraction  theorems (e.g., \cite{VW}), which requires the data be independently distributed.
Nevertheless,  the  idea of using covariates  would still be applicable for serial dependent data.
For instance, it is not difficult to allow for weak serial correlations when the data are not heavy-tailed, by using
the sieve least squares estimator $\widetilde\bSigma$ (introduced in Section \ref{Sieve_LS})  in place of the Huber's estimator $\widehat{\bSigma}$.  
We conduct numerical studies when the data are serially correlated  in  the simulations, and find that the proposed methods continue to perform well in the presence of serial correlations.



\subsection{Rates of convergence}

 We present the rates of convergence in the following theorems, and discuss the statistical insights in the next subsection.
Recall $\widehat{\bLambda}=(\widehat{\blambda}_i: i\leq N)$.
\begin{thm}[Loadings] \label{t31}
Under Assumptions \ref{a1}--\ref{asn},  there is an invertible matrix $\bH$,  as $ T, J\to\infty$, and $N$ either grows or stays constant,
  \begin{eqnarray}
\frac{1}{N}\sum_{i=1}^N\|\widehat{\blambda}_i-\bH'\blambda_i\|^2&=&O_P\left(\frac{J}{T}+\frac{1}{J^{2\eta-1}}\right)\chi_N^{-1},\label{e3.3}\\
\max_{i\leq N}\|\widehat{\blambda}_i-\bH'\blambda_i\|&=&O_P\left( \sqrt{\frac{J\log N}{T}}+\frac{1}{J^{\eta-1/2}}\right)\chi_N^{-1/2}.  \cr
\end{eqnarray}

\end{thm}
The  optimal rate for $J$ in  (\ref{e3.3}) is $J\asymp T^{1/(2\eta)}$, which results in
\begin{equation}\label{e4.4}\frac{1}{N}\sum_{i=1}^N\|\widehat{\blambda}_i-\bH'\blambda_i\|^2= O_P(T^{-(1- \frac{1}{2\eta})}\chi_N^{-1}).
\end{equation}
Here    $\eta$ represents the smoothness of $E(\bff_t|\bx_t=\cdot)$, as defined in Assumption \ref{a3.3}.  
 

 Define
$$
J^{*}= \min\left\{   (TN)^{1/(2\eta)}, ( \frac{T}{\log N})^{1/(1+\eta)} \right\}.
$$

\begin{thm}[Factors] \label{t32} Let   $J\asymp J^{*}$. Suppose $(J^*)^3\log^2N=O(T)$,   and Assumptions \ref{a1}--\ref{asn} hold.
 For  $\bH$ in Theorem \ref{t31},  as $ T\to\infty$, and   $N$ either grows or stays constant, we have
 \begin{equation*}
  \frac{1}{T}\sum_{t=1}^T\|\widehat\bg(\bx_t)-\bH^{-1}\bg(\bx_t)\|^2=
  O_P\left(  r_{T, N}^* + (\frac{\log N}{T})^{2-\frac{3}{1+\eta}} \right),
 \end{equation*}
 where $r_{T, N}^* = \frac{J^{*2}}{T^2}\chi_N^{-1}
  + {\frac{J^*\|\cov(\bgamma_t)\|}{T}}+ (\frac{1}{TN})^{1-\frac{1}{2\eta}}$ and
   \begin{eqnarray} \label{e4.5}
  \frac{1}{T}\sum_{t=1}^T\|    \widehat{\bgamma}_t-\bH^{-1}\bgamma_t
 \|^2&=&O_P\left(  r_{T, N}^*  + (\frac{\log N}{T})^{2-\frac{4}{1+\eta}}  \right)\chi_N^{-1}\cr
 &&+O_P\left(\frac{1}{N}\right) .
 \end{eqnarray}  \end{thm}
These two convergences  imply the rate of convergence of the estimated factors due to $\widehat\bff_t=\widehat \bg(\bx_t)+\widehat{\bgamma}_t$.

 \begin{rem}
For a general $J$, the rates of convergence of the two factor components are
   \begin{equation}\label{e3.8}
 \frac{1}{T}\sum_{t=1}^T\|\widehat\bg(\bx_t)-\bH^{-1}\bg(\bx_t)\|^2= O_P\left(  r_{T, N} +\frac{J^3\log^2 N}{T^2} \right),
\end{equation}
where $r_{T, N} = \frac{J^2}{T^2}\chi_N^{-1}   +\frac{J\| \cov(\bgamma_s)\|}{T}+J^{1-2\eta}+\frac{J}{TN}$ and
   \begin{eqnarray}\label{e4.7ad}
\qquad \qquad \frac{1}{T}\sum_{t=1}^T\|    \widehat{\bgamma}_t-\bH^{-1}\bgamma_t
\|^2  =   O_P\left(  r_{T, N} + \frac{ J^4\log^2 N }{T^2} \right) \chi_N^{-1} + O_P\left(\frac{1}{N}\right).
\end{eqnarray}
In fact   $J\asymp J^*$  is  the optimal choice  in (\ref{e3.8})  ignoring the terms involving $\|\cov(\bgamma_s)\|$ and $\chi_N$. The convergence rates presented in Theorem \ref{t32} are obtained from  (\ref{e3.8}) and (\ref{e4.7ad}) with this choice of  $J$.

The presented rates  connect well with the literature on both standard  nonparametric   sieve estimations and the high-dimensional factor models. To illustrate this, we discuss in more detail about the rate of convergence in (\ref{e3.8}).  This rate is given by:
$$
O_P\left(  \underbrace{\frac{J^2}{T^2}\chi_N^{-1}}_{\substack{\text{effect of} \\ \text{estimating $\bLambda$}}}   +\underbrace{\frac{J\| \cov(\bgamma_s)\|}{T}+\frac{J}{TN}+J^{1-2\eta}}_{\substack{\text{nonparametric sieve}\\ \text{estimation error}}} +\underbrace{\frac{J^3\log^2 N}{T^2}}_{\substack{\text{higher order from} \\ \text{Huber's M-estimation}}} \right).
$$
More specifically, we have, for $\be_t =\bLambda\bgamma_t+\bu_t$,
\begin{equation} \label{e4.8ad}
\by_t=\bLambda\bg(\bx_t)  +\be_t,\quad E(\be_t|\bx_t)=0.
\end{equation}
If $\bLambda$ were known, we would  estimate   $\bg(\cdot)$    by regressing the estimated  $E(\by_t|\bx_t)$ on $ \bLambda$.  Then standard nonparametric results  show that  the rate of convergence  in this ``oracle sieve regression'' (knowing $\bLambda$) would be
$$
\frac{J\| \cov(\bgamma_s)\|}{T}+\frac{J}{TN} +J^{1-2\eta}.
$$
   As we do not observe $\bLambda$, we are running the regression (\ref{e4.8ad}) with $\widehat{\bLambda}$ in place of $\bLambda$. This leads to an additional term $\frac{J^2}{T^2}\chi_N^{-1}$ representing the effect of estimating $\bLambda$, which also depends on the strength of the signal $\chi_N$.
Finally,  Huber's M-estimation to estimate $E(\by_t|\bx_t)$  gives rise to a higher order term $\frac{J^3\log^2 N}{T^2}$, and is often negligible.


\end{rem}



 \subsection{ The signal-noise regimes}\label{sec:sigal_noise}
We see that the rates   depend on $\cov(\bgamma_t)$ and $\chi_N$.
  Because
$E\bff_t\bff_t'=\bSigma_{f|x}+\cov(\bgamma_t)$, they  are related through
\begin{equation}\label{e4.7}
c\leq  \chi_N+\|\cov(\bgamma_t)\|\leq C_1
\end{equation}
for some $c, C_1 >0$,  assuming that there is $c>0$ so that $\|E\bff_t\bff_t'\|>c$.
 For comparison, we state the rates of convergence of the benchmark PCA estimators: (e.g., \cite{SW02, bai03}) there is a rotation matrix $\tilde\bH$, so that the PCA estimators $( \widetilde\blambda_i, \widetilde\bff_t)$ satisfy:
  \begin{equation} \label{e4.8}
   \frac{1}{N}\sum_{i=1}^N\|    \widetilde\blambda_i-\tilde\bH'\blambda_i
\|^2= O_P( \frac{1}{T}+\frac{1}{N}),\qquad \frac{1}{T}\sum_{t=1}^T\|    \widetilde\bff_t-\tilde\bH^{-1}\bff_t
\|^2= O_P( \frac{1}{T}+\frac{1}{N}).
\end{equation}

 The first interesting phenomena we observe is that both the estimated loadings and  $\bg(\bx_t)$ are consistent even if $N$ is finite, due to the ``exact identification". In contrast, the PCA estimators requires a growing $N$.
For more detailed comparisons, we consider  three regimes  based on the explanatory power of the factors using $\bx_t$.
To simplify our discussions, we  consider  the rate-optimal  choices of $J$,   and   ignore the   sieve approximation errors, so $\eta$ is treated  sufficiently large.

 \textit{Regime I: strong explanatory power: $\|\cov(\bgamma_t)\|\to 0.$} Because of (\ref{e4.7}), $\chi_N$ is bounded away from zero.   In this case, (\ref{e4.4})-(\ref{e4.5}) approximately imply  (for sufficiently large $\eta$):
  \begin{eqnarray*}
  \frac{1}{N}\sum_{i=1}^N\|\widehat{\blambda}_i-\bH'\blambda_i\|^2&=& O_P\left(\frac{1}{T}\right),\cr
 \frac{1}{T}\sum_{t=1}^T\|\widehat\bg(\bx_t)-\bH^{-1}\bg(\bx_t)\|^2&=&O_P\left(  \frac{\|\cov(\bgamma_t)\|}{T}+
  \frac{1}{TN}+ (\frac{\log N}{T})^{2 } \right),
\cr
 \frac{1}{T}\sum_{t=1}^T\|    \widehat\bff_t-\bH^{-1}\bff_t
\|^2&=&O_P\left(   \frac{\|\cov(\bgamma_t)\|}{T}+ \frac{1}{N} +   (\frac{\log N}{T})^{2 } \right) .
\end{eqnarray*}
Compared to the rates of the  usual PCA estimators in  (\ref{e4.8}),   either the  new estimated   loadings (when $N=o(T)$)  or the new estimated factors (when $T=o(N)$) have  a faster rate of convergence.
 Moreover, if $\|\cov(\bgamma_t)\|= o((TN)^{-1}+T^{-2}\log^2N)$, then
  $\widehat\bg(\bx_t)$ directly estimates the latent factor at a very    fast rate of convergence:
  $$
   \frac{1}{T}\sum_{t=1}^T\|\widehat\bg(\bx_t)-\bH^{-1}\bff_t\|^2=O_P\left(
  \frac{1}{TN}+ (\frac{\log N}{T})^{2 } \right).
  $$
   The improved rates are reasonable due to the   strong explanatory powers from the covariates.

\textit{Regime II: mild explanatory power:  $\|\cov(\bgamma_t)\| $ is   bounded away from zero; $\chi_N$ is either bounded away from zero or decays slower than $ \frac{N}{T}$ in the case $N=o(T)$.}     In this regime, $\bx_t$ partially explains the factors, yet the unexplainable components are not negligible.
 (\ref{e4.4})-(\ref{e4.5}) approximately become:
  \begin{eqnarray}
  \frac{1}{N}\sum_{i=1}^N\|\widehat{\blambda}_i-\bH'\blambda_i\|^2&=& O_P\left(\frac{1}{T}  \chi_N^{-1}\right)\cr
 \frac{1}{T}\sum_{t=1}^T\|    \widehat\bff_t-\bH^{-1}\bff_t
\|^2&=&O_P\left(      \frac{ 1}{T} \chi_N^{-1}+  \frac{1}{N}\right) .
\end{eqnarray}
 We see that the  rate   for the estimated loadings is still faster than the  PCA  when $N$ is relatively small compared to $T$, while the rates  for the estimated factors are the same.  This is because
 $$
 \underbrace{\frac{1}{T}\chi_N^{-1}}_{\text{new rate for loadings}}\ll  \underbrace{\frac{1}{N}}_{\text{ PCA rate for loadings}}
 $$
  $$
 \underbrace{\frac{1}{T}\chi_N^{-1}+ \frac{1}{N}}_{\text{new rate for factors}}\asymp \underbrace{\frac{1}{N}}_{\text{ PCA rate for factors}}.
 $$
On one hand, due to the explanatory power from the covariates,   the  loadings  can be estimated well without having to consistently estimate the factors.
 On the other hand, as the covariates only partially explain the factors, we cannot improve rates of convergence in estimating the unexplainable components in the latent factors. However, since $\bgamma_t$ has smaller variability than $\bff_t$, it can still be better  estimated in terms of a  smaller  constant factor.

\textit{Regime III:  weak explanatory power: $\chi_N\to 0$ and  decays faster than $  \frac{N}{T}$ when $N\ll T$.}   In this case, we have
  \begin{eqnarray*}\frac{1}{N}\sum_{i=1}^N\|\widehat{\blambda}_i-\bH'\blambda_i\|^2=O_P(\frac{1}{T}\chi_N^{-1}) =\frac{1}{T}\sum_{t=1}^T\|    \widehat\bff_t-\bH^{-1}\bff_t
\|^2
\end{eqnarray*}
While the  new estimators   are still consistent, they perform worse than PCA. This finding is still reasonable because the signal is  so weak that    the conditional expectation  $E(\by_t|\bx_t)$ loses useful information  of the factors/loadings.   Consequently, estimation efficiency is lost when running PCA on the estimated covariance  $ E\{  E(\by_t|\bx_t) E(\by_t|\bx_t)' \}$.

In summary,  improved rates of convergence can be achieved so long as the covariates can (partially) explain the latent factors, this corresponds to     either the mild or the strong explanatory power case.  The degree of improvements depend on the strength of the signals. In particular, the consistent estimation for factor loadings can  also be achieved even under finite $N$.  On the other hand, when the explanatory power is too weak, the rates of convergence would be slower than those of the benchmark estimator.

\section{Testing the Explanatory Power of Covariates}\label{sec:Test}

  We aim to test: (recall that $\bgamma_t=\bff_t-E(\bff_t|\bx_t)$)
\begin{equation}\label{e4.1add}
H_0: \cov(\bgamma_t)=0.
\end{equation}
Under $H_0$, $\bff_t=E(\bff_t|\bx_t)$ over the entire sampling period $t=1,..., T$,  implying that observed covariates  $\bx_t$ fully explain the true factors $\bff_t$.    In empirical applications with ``observed factors", what have been often used   are in fact $\bx_t$. Hence our proposed test can be applied to empirically validate the explanatory power of these ``observed factors".

The Fama-French three-factor model  \citep{FF}  is one of the most celebrated  ones in empirical asset pricing.
 They modeled the excess return $r_{it}$ on security or portfolio $i$ for period $t$ as
 $$
 r_{it} =\alpha_i +b_i r_{Mt}  +  s_i \SMB_t +h_i \HML_t  +  u_{it},
 $$ where  $r_{Mt}, \SMB_t$ and $\HML_t$   respectively represent the the excess returns of the market,  the difference of returns between stocks with small and big market capitalizations (``small minus big"),  and  the difference of returns between stocks with high book to equity ratios and those with low book to equity ratios (``high minus low").  Ever since its proposal,  there is much evidence that the three-factor model can leave the cross-section of expected stock returns unexplained. Different factor definitions have been explored, e.g., \cite{carhart1997persistence} and  \cite{novy2013other}.
   \cite{fama2015five}  added profitability and investment factors to the
 three-factor  model.  They conducted   GRS tests   \citep{GRS} on the five-factor models and its different variations. Their tests   ``reject all models as a complete description of expected returns".




On the other hand,  the Fama-French factors, though imperfect,     are  good proxies for the true unknown factors.
Consequently,   they form a natural choice for $\bx_t$. These observables are actually diversified portfolios, which have explanatory power on the latent factors $\bff_t$, as supported by financial economic theories as well as empirical studies.
The test proposed in this validates the specification of these common covariates as ``factors".



\subsection{The Test Statistic}

Our  test is based on a Wald-type weighted  quadratic statistic
$$
S(\bW):=\frac{N}{T}\sum_{t=1}^T\widehat{\bgamma}_t'\bW\widehat{\bgamma}_t=\frac{1}{TN}\sum_{t=1}^T(\by_t-\widehat E(\by_t|\bx_t))'\widehat{\bLambda}\bW\widehat{\bLambda}'(\by_t-\widehat E(\by_t|\bx_t)).
$$
The weight matrix normalizes the test statistic, taken as $\bW=$ AVar$(\sqrt{N}\widehat{\bgamma}_t)^{-1}$, where  AVar$(\widehat{\bgamma}_t)$ represents the asymptotic   covariance matrix of $\widehat{\bgamma}_t$ under the null, and is given by
$$
\text{AVar}(\sqrt{N}\widehat{\bgamma}_t)=\frac{1}{N}\bH' \bLambda'\bSigma_u\bLambda\bH.
$$
As $\bSigma_u$ is a high-dimensional covariance matrix, to  simplify the technical arguments, in this section we assume $\{u_{it}\}$ to be cross-sectionally uncorrelated, and estimate $\bSigma_u$ by:
$$
\widehat{\bSigma}_u=\diag\{\frac{1}{T}\sum_{t=1}^T\widehat u_{it}^2, i=1,..., N\}, \quad \widehat u_{it}=y_{it}-\widehat{\blambda}_i'\widehat\bff_t.
$$
The feasible test statistic is defined as
$$
S := S(\widehat\bW),\quad \widehat\bW:=(\frac{1}{N}\widehat{\bLambda}'\widehat{\bSigma}_u\widehat{\bLambda})^{-1}.
$$
 We reject the null hypothesis for large values of $S$.
It is straightforward to allow     $\bSigma_u$ to be a non-diagonal but a sparse covariance,  and proceed   as in \cite{Bickel08a}.    We expect  the asymptotic analysis to be quite involved, and do not pursue it in this paper.





We  show that the test statistic    has the following asymptotic expansion: 
$$
 S= \bar S+ o_P(\frac{1}{\sqrt{T}}),
$$
where $$
\bar S=\frac{1}{T}\sum_{t=1}^T \bu_t'\bLambda (\bLambda'\bSigma_u\bLambda)^{-1}\bLambda'\bu_t.
$$
Thus the limiting distribution is determined by that of $\bar S$.  Note that a  cross-sectional central limit theorem  implies, as $N\to\infty$,
$$
(\frac{1}{N}\bLambda'\bSigma_u\bLambda)^{-1/2}\frac{1}{\sqrt{N}}\bu_t'\bLambda\to^d\mathcal{N}(0,\bI_K).
$$
 Hence each component of $\bar S$ can be roughly understood as    $\chi^2$-distributed with degrees of freedom $K$   being the number of common factors, whose variance is $2K$.  This motivates the following assumption.

 \begin{ass}\label{a4.1}
 Suppose
 $
\frac{ 1}{T}\sum_{t=1}^T \var(\bu_t'\bLambda (\bLambda'\bSigma_u\bLambda)^{-1}\bLambda'\bu_t  )\to2K$  as $T, N\to\infty$.

 \end{ass}

 We now state the null distribution in the following theorem.

 \begin{thm}\label{t41}
 Suppose
 $\{u_{it}\}_{i\leq N}$ is cross-sectionally independent, and  Assumption \ref{a4.1} and assumptions of Theorem \ref{t32} hold. Then, when  $J^4N\log N =o(T^{3/2})$, $T=o(N^2)$, $N\sqrt{T}=o(J^{2\eta-1})$,  as    $T, N\to\infty$,
$$\sqrt{\frac{T}{2K}} (S-K)\to^d\mathcal{N}(0,1).
$$

 \end{thm}



\subsection{Testing market risk factors for S\&P 500 returns}
\numberwithin{table}{section}
\numberwithin{figure}{section}

We     test the explanatory power of the observable proxies for the true factors using  S\&P 500 returns.
For each given group of observable proxies, we set the number of common factors $K$ equals the number of observable proxies.
We calculate the  excess returns for the stocks in S\&P 500 index that are  collected from CRSP.   
We consider three groups of proxy factors ($\bx_t$) with increasing information: (1) Fama-French 3 factors (FF3);
(2) Fama-French 5 factors (FF5); and (3) Fama-French 5 factors plus 9 sector SPDR ETF's (FF5+ETF9).
Here the sector SPDR ETF's, which are intended to  track the 9 largest S\&P sectors.
The detailed descriptions of sector SPDR ETF's are listed in Table \ref{E2_tab_ETF}.

\begin{table}[htbp]
\begin{center}
\small
\caption{Sector SPDR ETF's (data available from Yahoo finance)}
\label{E2_tab_ETF}
\begin{tabular}{c|c|c|c|c|c}
\hline\hline
Code		  &  Sector & Code		  &  Sector & Code		  &  Sector
\\\hline
XLE &  Energy & XLB &  Materials & XLI &  Industrials
\\
XLY &  Consumer discretionary & XLP &  Consumer staples
& XLV &  Health care
\\
XLF &  Financial & XLK &  Information technology
& XLU &  Utilities
\\\hline\hline
\end{tabular}
\end{center}
\end{table}
 
\begin{figure}[htbp]
 \centering
 \includegraphics[width=4 in]{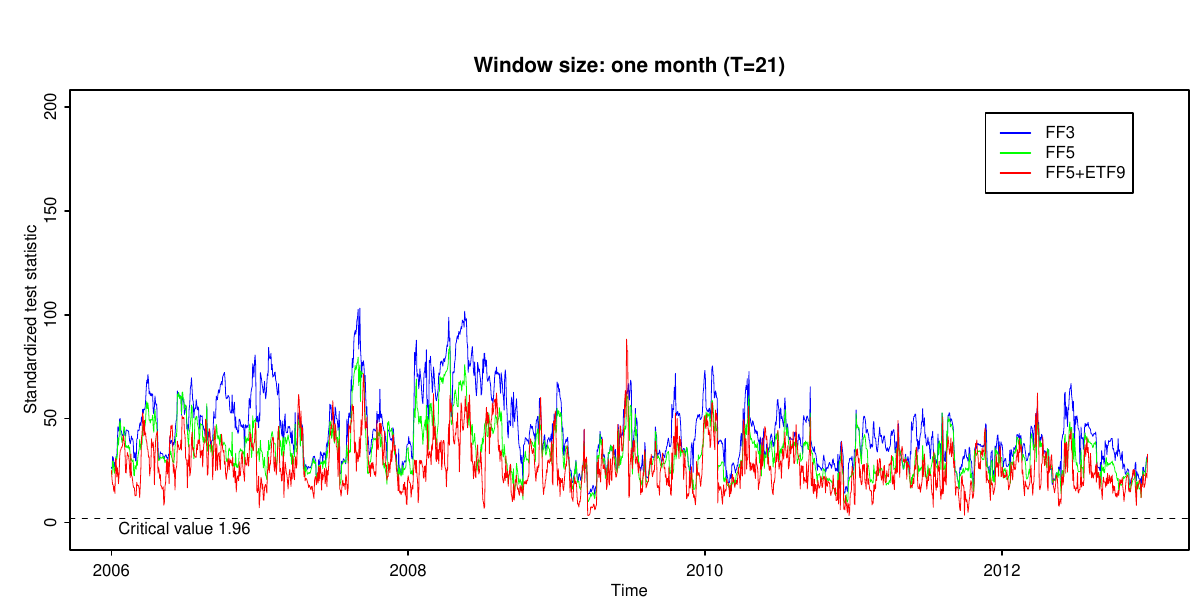}
 \includegraphics[width=4 in]{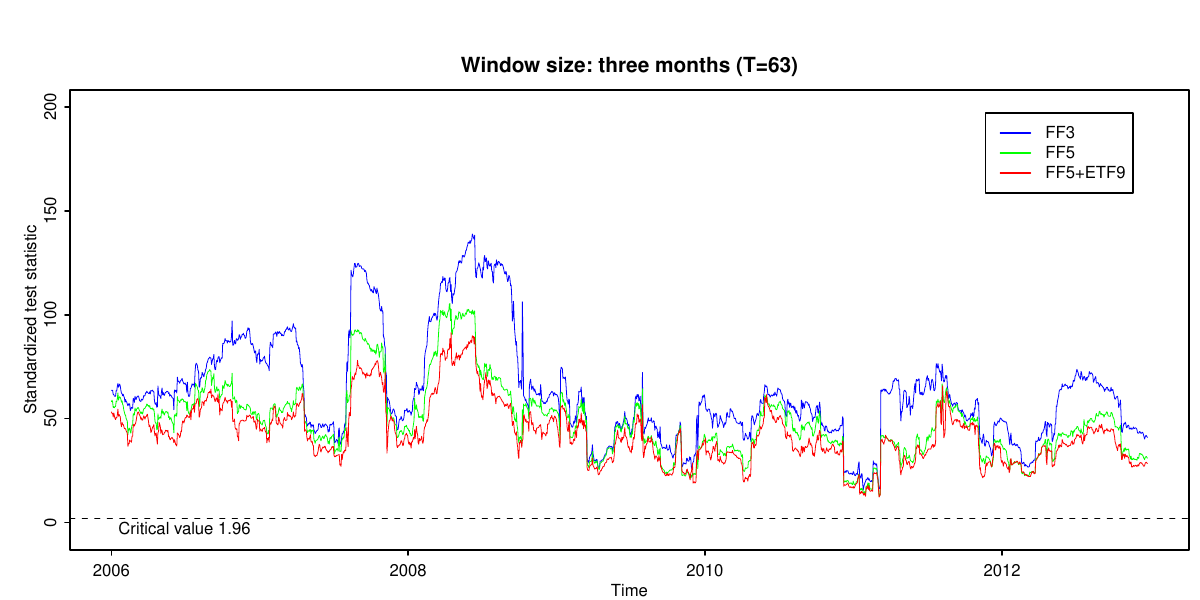}
 \includegraphics[width=4 in]{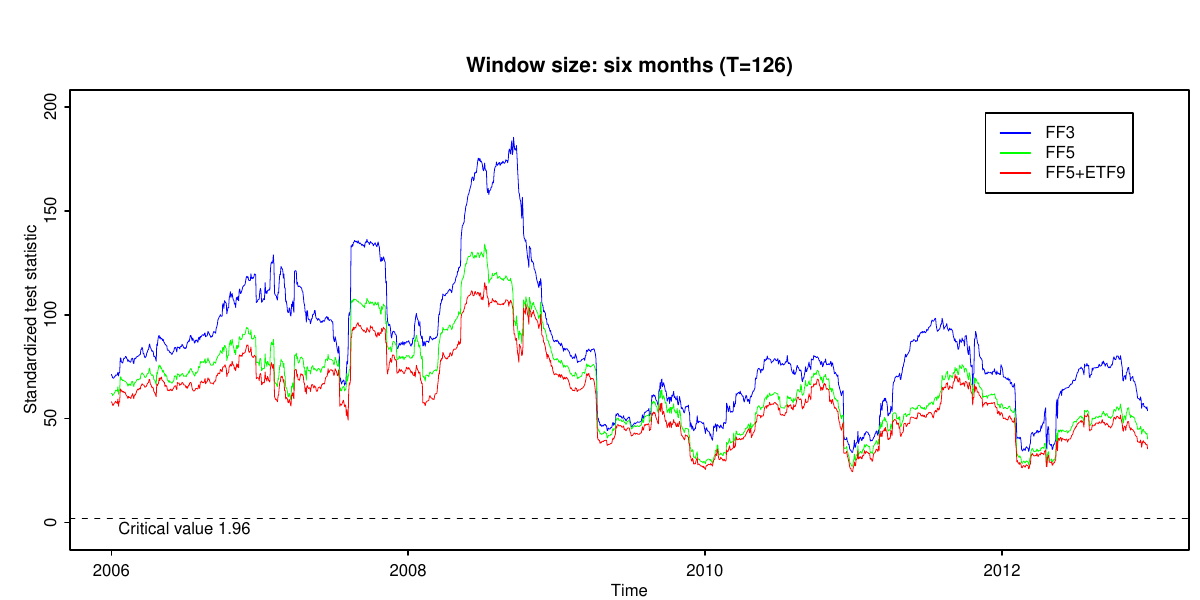}
 \caption{S\&P 500 daily returns: plots for standardized test statistic $S $ for various window sizes. The dotted line is critical value 1.96.}
 \label{E2_fig_test_1}
\end{figure}

\begin{figure}[htbp]
 \centering
 \includegraphics[width=4 in]{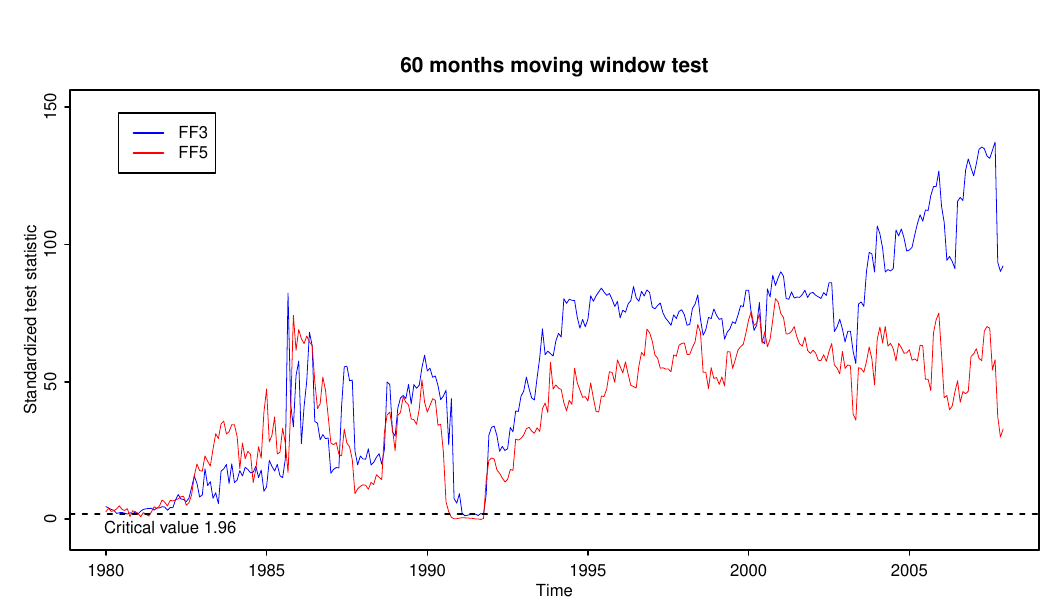}
 \includegraphics[width=4 in]{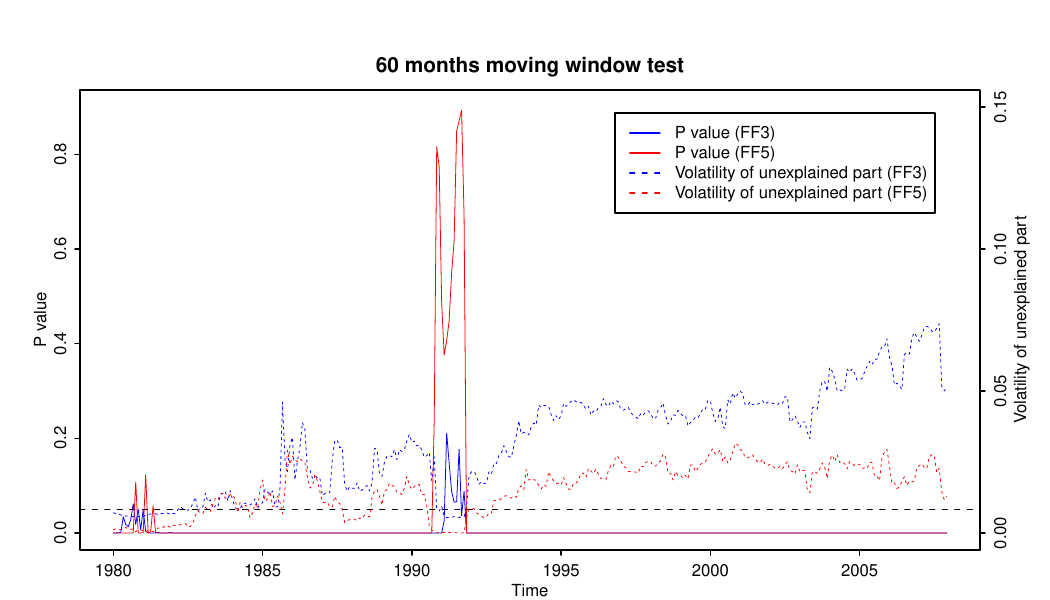}
 \caption{S\&P 500 monthly returns: plots for standardized test statistic $S $, P-value and the volatility of the part of factors that can not be explained by the proxy factors.}
 \label{E2_fig_test_2}
\end{figure}

We consider tests using both daily and monthly data. For the daily data,  
we collect  393 stocks   that have complete daily closing prices from
January 2005 to December 2013, with a time span of 2265 trading days. 
 We apply moving window tests with the window size ($T$) equals one month, three months or six months. The testing window moves one trading day forward per test. Within each testing window, we calculate the standardized test statistic $S$ for three groups of proxy factors.

As for the monthly excess returns, we use stocks   that have complete record from
January 1980 to December 2012, which contains 202 stocks with a time span of 396 months.
Here we only consider the first two groups of proxy factors as sector SPDR ETF's are introduced since 1998.
 The window size equals sixty months and moves one month forward per test. Within each testing window, besides standardized test statistic and p-value, we also estimate the volatility of $\bgamma_t$,  the part of factors that can not be explained by $\bx_t$ as: $$
\hat{\mbox{Vol}}(\bgamma_t)=\frac{1}{21T}\sum\limits_{t=1}^T\widehat\bgamma_t'\widehat\bgamma_t,
$$
where  there are  21 trading days per month. The sieve basis is chosen as the additive Fourier basis with $J=5$.
We set the tuning parameter $\alpha_T=C\sqrt{\frac{T}{\log(NJ)}}$ with constant $C$ selected by the 5-fold cross validation.

 For the daily data, the  plots of $S$ under various scenarios are reported in Figure \ref{E2_fig_test_1}. Under all scenarios, the null hypothesis ($H_0: \cov(\bgamma_t)=0$) is rejected as  $S $ is always  larger than the critical value 1.96. This suggests a strong evidence that the proxy factors can not fully explain the estimated common factors.
Under all window sizes, a larger group of proxy factors tends to yield smaller statistics, demonstrating  stronger explanatory power for estimated common factors. Also, we find the test statistics increase while the window size increases.  

The results for the monthly data are   reported in Figure \ref{E2_fig_test_2}. For both Fama-French 3 factors and 5 factors,
the null hypothesis is rejected most of the time except in early 1980s and 1990s. When the null hypothesis is accepted, Fama-French 5 factors tend to yield larger p-values. The estimated volatility of unexplained part are close to zero over these two periods.  For the rest of the time, the standardized test statistics are much larger than the critical value 1.96 and hence the p-values are close to zero. Also the estimated volatilities are not close to zero.  This indicates the proxy factors can not fully explain the estimated common factors during these testing periods.

\medskip


\section{Forecast the excess return of US government bonds}\label{sec:Emp1}

We apply our method to forecast the excess return of U.S. government bonds.
The bond excess return  is the one-year bond return in excess of the risk-free rate. To be more specific,
we buy an $n$ year bond, sell it as an $n-1$ year bond in the next year and excess the one-year bond yield as the risk-free rate.
Let $p_t^{(n)}$ be the log price of an $n$-year discount bond at time $t$.
Denote $\zeta_t^{(n)} \equiv -\frac{1}{n}p_t^{(n)}$ as the log yield with $n$ year maturity,
and $r_{t+1}^{(n)} \equiv p_{t+1}^{(n-1)}-p_t^{(n)}$ as the log holding period return.
The goal of one-step-ahead forecast  is to forecast $z_{T+1}^{(n)}$, the   excess return with maturity of $n$ years in period $T+1$, where
$$ z_{t+1}^{(n)}=r_{t+1}^{(n)}-\zeta_t^{(1)},\quad t=1, \ \cdots \ , T.$$

For a long time, the literature has found a significant predictive power of the excess returns of U.S. government bonds. For instance, \cite{ludvigson2009macro, ludvigson2010factor} predicted the bond excess returns with observable variables based on a factor model using 131 (disaggregated) macroeconomics variables.
They achieved the out-of-sample  $R^2 \approx 21\%$ when forecasting one year excess bond return with maturity of two years.
Using the proposed   method, this  section develops a new way of incorporating the explanatory power of the observed characteristics, and investigates
the  robustness of the conclusions in existing literature.

We analyze monthly data spanned from January 1964 to December 2003, which is available from the Center for Research in Securities Prices (CRSP). The factors are estimated from a macroeconomic dataset consisting of 131 disaggregated macroeconomic time series  \citep{ludvigson2010factor}.   The  covariates  $\bx_t$ are 8 aggregated macro-economic time series, listed in Table \ref{E1_tab1}.

\begin{table}[htbp]
\begin{center}
\caption{Components of $\bx_t$}
\label{E1_tab1}
\begin{tabular}{c|c}
\hline\hline
$x_{1,t}$ &  Linear combination of five forward rates
\\\hline
$x_{2,t}$ &  Real gross domestic product (GDP)
\\\hline
$x_{3,t}$ &  Real category development index (CDI)
\\\hline
$x_{4,t}$ &  Non-agriculture employment
\\\hline
$x_{5,t}$ &  Real industrial production
\\\hline
$x_{6,t}$ &  Real manufacturing and trade sales
\\\hline
$x_{7,t}$ &  Real personal income less transfer
\\\hline
$x_{8,t}$ &  Consumer price index (CPI)
\\\hline
\end{tabular}
\end{center}
\end{table}

\subsection{Heavy-tailed data and robust estimations}
\begin{figure}[htbp]
 \centering
 \includegraphics[width=5 in]{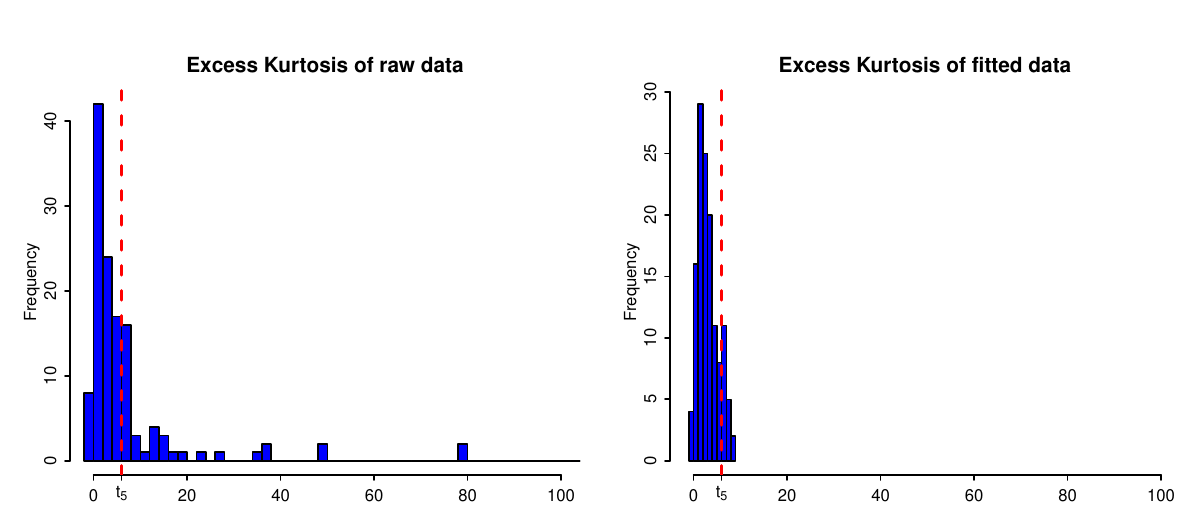}
 \caption{Excess kurtosis of the macroeconomic panel data.
 Left panel shows 43 among 131 series in the raw data are heavy tailed. Right panel shows the robustly fitted data $\hat{E}(\by_t|\bx_t)$ are no longer severely heavy-tailed.}
 \label{E1_fig1}
\begin{center}
\end{center}
\end{figure}

We first examine  the excess kurtosis for the time series to assess the tail distributions.  The left panel of Figure \ref{E1_fig1} shows 43 among the 131 series have   excess kurtosis greater than 6. This indicates the tails of their distributions are fatter than the $t$-distribution with degrees of freedom 5.  On the other hand, the right panel of Figure \ref{E1_fig1} reports the histograms of excess kurtosis of the
``fitted data" $\hat{E}(\by_t|\bx_t)$ (the robust estimator of  $E(\by_t|\bx_t)$ using Huber loss), which demonstrates that most series in the fitted data are no longer severely heavy-tailed.   

The tuning parameter in the Huber loss is of order $\alpha_{T}=C_\alpha \sqrt{\frac{T}{\log(NT)}}$. In this study, the constant $C_{\alpha}$ and the degree of sieve approximation $J$ are selected by the out-of-sample 5-fold cross validation as described in Section \ref{sec_tunning}.



\subsection{Forecast results}\label{sec:Emp1.2}

We denote our proposed method by  SPCA (\textit{smoothed PCA}),  and compare it with  SPCA-LS (which uses $\widetilde\bSigma$, the   least-squares based smoothed PCA,  described in Section \ref{Sieve_LS}) and the benchmark PCA.
 We conduct  one-month-ahead out-of-sample forecast of the bond excess returns. The forecast uses the information in the past 240 months,  starting from January 1984 and rolling forward to December 2003.   We   compare three approaches to estimating the factors:  SPCA, SPCA-LS,  and  the usual PCA. Also we consider two forecast models as follows:
\begin{eqnarray}
& \text{Linear model:} \quad & z_{t+1}=\alpha + \bbeta ' \bW_t+\epsilon_{t+1}, \label{Linear-model}
\\
& \text{Multi-index model:}  \quad & z_{t+1}= h({\bfsym {\psi}}_1'  \bW_t,\  \cdots , \ {\bfsym {\psi}}_L'  \bW_t)+\epsilon_{t+1}, \label{multi-indx}
\end{eqnarray}
where $\alpha$ is the intercept and $h$ is a nonparametric function. 
The covariate $\bW_t$ is either  $\bff_t$ or an augmented vector  $(\bff_t',\bx_t')'$.  Here, the latent factors $\bff_t$ are used by the three methods mentioned above in order to compare their effectiveness. The  multi-index model allows more general nonlinear forecasts and are estimated by using the sliced inverse regression \citep{li1991sliced}.
The number of  indices $L$ is estimated by the ratio-based method suggested in \cite{LamYao} and  is usually 2 or 3. We approximate $h$ using a weighted additive model $h({\bfsym {\psi}}_1'  \bW_t,\  \cdots , \ {\bfsym {\psi}}_L'  \bW_t)=\sum_{l=1}^L  g_l({\bfsym {\psi}}_l'  \bW_t)$. Each individual nonparametric function $g_l(\cdot)$ is smoothed by the local linear approximation.  

The performance of each method is assessed by the out-of-sample $R^2$.
Let $\widehat z_{T+t+1|T+t}$ be the forecast of $z_{T+t+1}$ using the data of the previous $T$ months: $1+t,...,T+t$ for $T=240$ and  $t=0,..., 239$.
 The forecast performance is assessed by the out-of-sample $R^2$, defined as
$$
R^2=1- \frac{\sum\limits_{t=0}^{239}(z_{T+t+1}-\hat{z}_{T+t+1|T+t})^2}{\sum\limits_{t=0}^{239}(z_{T+t+1}-\bar{z}_t)^2}, 
$$
where $\bar{z}_t$ is the sample mean of $z_t$ over the sample period $[1+t, T+t]$.
The $R^2$ of various methods are reported in Table  \ref{E1_tab3}.  We notice that factors estimated by  SPCA and SPCA-LS can explain more variations in bond excess returns with all maturities than the ones estimated by PCA. 
SPCA yields a 44.6\% out-of-sample $R^2$ for forecasting the bond excess returns with two year maturity, which is much higher than the best out-of-sample predictor found in \cite{ludvigson2009macro}.
It is also observed that  the forecast   based on either SPCA or SPCA-LS cannot be improved by adding any covariate in $\bx_t$.   We argue that, in this application, the information of $\bx_t$ should be mainly used as the explanatory power for   the factors.


We summarize the observed results in the following aspects:

\begin{enumerate}
\item The factors estimated using additional covariates  lead to significantly improved out-of-sample forecast on the  US bond excess returns compared to the ones estimated by PCA.

\item As many series in the panel data are heavy-tailed, the robust-version of our method (SPCA) can  result in improved out-of-sample forecasts.

\item The  multi-index models  yield  significantly larger out-of-sample $R^2$'s than those of the linear forecast models.

\item The observed covariates $\bx_t$ (e.g. forward rates, employment and inflation) contain strong explanatory powers of the latent factors.  The gain of forecasting bond excess returns is more substantial when these covariates are   incorporated to estimate the common factors (using the proposed procedure) than  directly used for forecasts.

\end{enumerate}

\begin{table}[htbp]
\footnotesize
\begin{center}
\caption{Forecast out-of-sample $R^2$ (\%): the larger the better.    }
\label{E1_tab3}
\begin{tabular}{c|cccc|cccc|cccc}
\hline\hline
$\bW_t$               & \multicolumn{4}{c|}{SPCA }          & \multicolumn{4}{c|}{SPCA-LS }
                      & \multicolumn{4}{c}{PCA}
\\\hline
                      & \multicolumn{4}{c|}{Maturity(Year) }        & \multicolumn{4}{c|}{Maturity(Year) }
                      & \multicolumn{4}{c}{Maturity(Year) }
\\
                      & 2   & 3   & 4   & 5                   & 2   & 3   & 4   & 5
                      & 2   & 3   & 4   & 5
\\\hline
                           & \multicolumn{4}{c|}{ }        & \multicolumn{4}{c|}{linear model}
                      & \multicolumn{4}{c}{ }   
                      \\
$\bff_t$              & 38.0 & 32.7 & 25.6 & 22.9   & 37.4 & 33.4 & 25.4 & 22.6
                      & 23.0 & 20.7 & 16.8 & 16.5
\\
$(\bff_t',\ \bx_t')'$ & 37.7 & 32.4 & 25.4 & 22.7   & 37.1 & 31.9 & 25.3 & 22.1
                      & 23.9 & 21.4 & 17.4 & 17.5
                      \\
                              & \multicolumn{4}{c|}{ }        & \multicolumn{4}{c|}{multi-index model }
                      & \multicolumn{4}{c}{ } 
                         \\
                      $\bff_t$              & 44.6 & 43.0 & 38.8 & 37.3   & 41.2 & 39.1 & 35.2 & 34.1
                      & 30.1 & 25.5 & 23.2 & 21.3
\\
$(\bff_t',\ \bx_t')'$ & 41.5 & 38.7 & 35.2 & 33.8   & 41.1 & 35.7 & 32.2 & 30.0
                      & 30.8 & 26.3 & 24.6 & 22.0
\\\hline
\end{tabular}
\end{center}
\end{table}

\section{Simulation Studies}\label{sec:Sim}

\subsection{Model settings}\label{sec:sim1}
We use simulated examples to demonstrate the finite sample performance of the proposed  method, which is denoted by SPCA (\textit{smoothed PCA}),  and compare it with  SPCA-LS (which uses $\widetilde\bSigma$, the   least-squares based smoothed PCA,  described in Section \ref{Sieve_LS}) and the benchmark PCA.
We set  $N=40, \ T=100$ and $K=5$.  The supplementary material contains additional simulation results under other $N$ and $T$ combinations,  as well as the case of serially dependent data. The findings are similar.

\medskip

Consider the following data generating process,
\begin{eqnarray*}\label{sim_model}
\by_t& =& \bLambda \bff_t+\bu_t,
\quad \text{and} \quad
\bff_t =\tilde{\sigma}(g)\bg^0(\bx_t)+  \tilde{\sigma}(\gamma){\bgamma}^0 _t,   \quad t=1, \ \cdots \ , T,
\end{eqnarray*}
where  $\bLambda$ is drawn from i.i.d. standard Normal distribution and   $\bu_t$ is drawn from either the i.i.d standard Normal distribution  or i.i.d. re-scaled Log-Normal distribution $c_1 \{ \exp(1+1.2 \zeta ) - c_2\}$, where $\zeta \sim \mathcal{N}(0, 1)$ and $c_1, c_2 >0$ are chosen such that $u_{it}$ has mean zero and variance $1$.

Here $\tilde{\sigma}(g)$ and $\tilde{\sigma}(\gamma)$ respectively represent the signal and noise levels. Set   $\tilde{\sigma}(g)^2+\tilde{\sigma}(\gamma)^2=1$ and $\tilde{\sigma}(g)^2/\tilde{\sigma}(\gamma)^2=\omega$, where $\omega$ controls the ratio between the explained and unexplained parts in the latent factors.
To address different signal-noise regimes, we set $\omega=10, 1$ and $0.1$ to represent strong, mild and weak explanatory powers respectively.
 The baseline  $\bgamma_t^0$ is drawn from i.i.d. standard Normal distribution and the baseline  function $\bg^0(\cdot)$ is set to be one of the following two models:
\begin{enumerate}
\item[(I)] {\sc Linear model}:\ We set $d=K$ and $\bx_t$ is drawn from i.i.d. standard Normal distribution.  Let $\bg^0(\bx_t)= \bD \bx_t$, where $\bD$ is a $K \times K$ matrix with each entry drawn
from $U[1,\ 2]$;
\item[(II)] {\sc Nonlinear model}:\ We set $d=1$ and $x_t$ is drawn from i.i.d. uniform distribution $[0,1]$. Let $\bg^0(x_t)= \{g_1^0(x_t), \cdots , g_K^0(x_t)\}'$ with $g_k^0(x_t)=a_k \cos(2 \pi k x_t) + b_k \sin(2 \pi k x_t)$ for $k=1, \cdots K$. The coefficients $a_k$ and $b_k$ are calibrated from a nonlinear test function $\theta(x)=\sin(x)+2\exp(-30x^2)$ with $x \in [0, 1]$ so that $\bg^0$ forms its leading Fourier bases.  To save the space, we refer to the example 2 of \cite{Dimatteo_01} for the plot of $\theta(x)$.

\end{enumerate}
For each $k\leq K$, we normalize $g_k^0(\bx_t)$ and $\gamma^0_{t,k}$ such that they have   means zero, and standard deviations one.

Throughout this section, the number of factors is estimated by the eigen-ratio method \citep{LamYao,AH}. In the following simulated examples, the eigen-ratio method can correctly select $K=5$ in most replications.  The sieve basis is chosen as the additive polynomial basis.  To account the scale of the noise variance, we also consider the tuning parameter in the Huber loss to admit $\alpha_{T, i}=C_\alpha\tilde{\sigma}_i\sqrt{\frac{T}{\log(NT)}}$, where $\tilde{\sigma}_i=\sqrt{ \frac{1}{T}\sum_t(y_{it}-\tilde E(y_{it}|\bx_t))^2}$ and $\tilde E(y_{it}|\bx_t)$ is smoothed by sieve least squares using additive polynomial basis of order 5.
In Subsection \ref{sec:sim2}, the tuning parameters $C_{\alpha}$ and $J$ are selected by the in-sample  5-fold cross validation, while in subsection \ref{sec:sim3}, they are chosen using the    out-of-sample 5-fold cross validation.

\subsection{In-sample Estimation}\label{sec:sim2}

First, we compare the in-sample model fitting   among SPCA, SPCA-LS and PCA under different scenarios. 
For each scenario, we conduct 200 replications.      As the factors and loading may be estimated up to a rotation matrix, the canonical correlations between the parameter and its estimator can be used to measure the estimation accuracy \citep{bai03}. 
For Model (I) and (II) we report the sample mean of the median of 5 canonical correlations between the true loading and factors  and the  estimated ones.

The results are presented in Table \ref{Ntab1_2_S}. SPCA-LS and SPCA are comparable for light-tail distributions, and are both slightly better than PCA.   This implies that we pay little price
for the  robustness and that the proposed estimators are potentially better than PCA when $N$ is relatively small, due to the  merit of the ``finite-$N$'' asymptotics of the proposed estimators.  However, when the error distributions have heavy tails, SPCA yields much better estimation   than other methods as expected.  SPCA-LS out-performs   PCA when $\bx_t$ has strong or mild explanatory powers of $\bff_t$ which is in line with the discussion in Section \ref{sec:sigal_noise}.  
 When $\omega = 0.1$, the observed $\bx_t$ is not as informative and hence the performance of SPCA and SPCA-LS are  close to regular PCA. 


\subsection{Out-of-sample Forecast}\label{sec:sim3}
We now consider using latent factors in a linear forecast model
$
z_{t+1} =\bbeta' \bff_t + \epsilon_{t+1},
$
where $\epsilon_{t}$ is drawn from i.i.d. standard normal distribution. For each simulation, the unknown coefficients in $\bbeta$ are independently drawn from uniform distribution $[0.5, \ 1.5]$ to cover a variety of model settings.

We conduct one-step ahead  rolling window forecast using the linear model by estimating $\bbeta$ and $\bff_t$. 
 The   factors are estimated from (\ref{sim_model}) by SPCA, SPCA-LS or PCA. In each replication, we generate $T+50$ observations in total. For $s=1, \ \cdots \ ,50$, we use the $T$ observations  $(z_s,..., z_{T+s-1})$
 to forecast $z_{T+s}$. We use PCA as the benchmark and define the relative mean squared error (RMSE) as:
\begin{equation*}
\RMSE=\frac { \sum\limits_{s=1}^{50}(\hat{z}_{T+s|T+s-1} - z_{T+s} )^2}
{\sum\limits_{s=1}^{50}(\tilde{z}_{T+s|T+s-1}^{PCA} - z_{T+s} )^2},
\end{equation*}
where $\hat{z}_{T+s|T+s-1} $ is the forecast of $z_{T+s}$ based on either SPCA or  SPCA-LS  while $\tilde{z}_{T+s|T+s-1}^{PCA}$ is the forecast based on PCA.
For each scenario, we simulate 200 replications and calculate the averaged RMSE as a measurement of  the one-step-ahead out-of-sample forecast.   

The results are presented in Table \ref{Ntab2_S}.  Again, when the tails of error distributions are light, SPCA and SPCA-LS perform comparably.  But SPCA outperforms SPCA-LS when the errors have heavy tails.  On the other hand, both SPCA and SPCA-LS outperform  PCA when $\bx_t$ exhibits strong or mild explanatory powers of $\bff_t$, but are slightly worse when $\omega$ is small. In general, the SPCA method performs the  best under heavy-tailed cases. 

\begin{table}[htbp]
\begin{center}
\small
\caption{{\bf Out-of-sample Forecast}: Mean RMSE of forecast when $N=40,T=100$: the smaller the better (with PCA as the benchmark)}
\label{Ntab2_S}
\begin{tabular}{c|c|cc|cc}
\hline\hline
&  &  \multicolumn{2}{c|}{Model (I)}
&  \multicolumn{2}{c}{Model (II)}
\\
  $\bu_t$ &  $\omega$               & {\scriptsize SPCA} & {\scriptsize SPCA-LS}  & {\scriptsize SPCA} & {\scriptsize SPCA-LS}

\\\hline
\multirow{3}{*}{Normal} & 10 & 0.86 & 0.85      & 0.88 & 0.87
\\
                             & 1  & 0.91 & 0.91      & 0.92 & 0.92
\\
                             & 0.1  & 1.01 & 1.01      & 1.02 & 1.01
\\\hline
\multirow{3}{*}{LogN}        & 10 & 0.45 & 0.60      & 0.49 & 0.64
\\
                             & 1  & 0.52 & 0.62      & 0.51 & 0.66
\\
                             & 0.1  & 0.55 & 0.65      & 0.56 & 0.70
\\\hline
\end{tabular}
\end{center}
\end{table}

\subsection{Compare with  the  interactive effect approach}
  Here we consider three pairs of sample sizes: $N=40,\ T=150$; $N=60,\ T=100$ and $N=60, \ T=150$. We compare the proposed SPCA method with SPCA-LS (Section \ref{Sieve_LS}), regular  PCA and pure least squares (LS), which models the covariates and estimates the parameters by simply using
$$
\min_{\bLambda,\{\bff_t\},\bbeta}\frac{1}{T}\sum_{t=1}^T\|\by_t-\bLambda \bff_t-\bx_t'\bbeta\|^2.
$$

In Tables \ref{Ntab1_2_S}--\ref{Supp_tab3}, we report sample mean of the median of 5 canonical correlations between the true loading and factors  and the  estimated ones. Under various sample size combinations, the findings are similar as discussed in Section \ref{sec:sim2}: (1) both SPCA and SPCA-LS outperform PCA under light-tail distributions when $\bx_t$ has strong or mild explanatory powers of $\bff_t$; (2)
when the error distributions have heavy tails, SPCA outperforms other methods as expected; (3)when $\bx_t$ has weak explanatory power, the performance of SPCA and SPCA-LS are  close to regular PCA; (4) under all simulated scenarios,  the LS approach gives the worst estimation performance.

\subsection{Serial dependent case}\label{serial_dep}
In this subsection, we compare the in-sample model fitting   among SPCA, SPCA-LS and PCA under serial dependences.
The simulation settings are similar as in Section 5.1 except both $\bx_t$ and $\bgamma_t$ are  generated from a stationary VAR(1) model as follows
\begin{align*}
\bx_t={\bf \Pi} \bx_{t-1} + \bepsilon_t, \quad
\bgamma_t={\bf \Pi} \bgamma_{t-1} + {\bf \eta}_t, \quad
t=1,\ \cdots, \ T,
\end{align*}
with $\bx_0={\bf 0}$ and $\bgamma_0= {\bf 0}$.
The $(i,  j)$th entry of ${\bf \Pi}$ is set to be 0.5 when $i=j$ and $0.4^{|i-j|}$ when $i \neq j$. In addition, $\bepsilon_t$ and ${\bf \eta}_t$ are drawn form i.i.d. $N({\bf 0}, \bI)$.

The performance under 200 replications are presented in Table \ref{Dtab1_2} below.  Our numerical findings for the independent data continue to hold for serially dependent data: both SPCA and SPCA-LS outperform PCA when  $\bx_t$ and $\bff_t$ are  serially correlated. SPCA gives the best performance when the error distributions are heavy-tailed.

\begin{table}[htbp]
\small
\begin{center}
\caption{{\bf In-sample Estimation}: Median of 5 canonical correlations of the estimated loadings/factors and the true
ones when $N=40,T=100$: the larger the better }
\label{Ntab1_2_S}
\begin{tabular}{c|c|c|ccc|ccc}
\hline\hline
& &  &  \multicolumn{3}{c|}{Model (I)}
&  \multicolumn{3}{c}{Model (II)}
\\
&   $\bu_t$ &  $\omega$
  & SPCA & SPCA-LS & PCA & SPCA & SPCA-LS & PCA
\\\hline
\multirow{6}{*}{Loadings} &
\multirow{3}{*}{Normal} & 10 & 0.91 & 0.91 & 0.82    & 0.90 & 0.90 & 0.75
\\
&                             & 1  & 0.88 & 0.89 & 0.82    & 0.84 & 0.84 & 0.75
\\
 &                            & 0.1  & 0.83 & 0.83 & 0.82    & 0.77 & 0.79 & 0.75
\\\cline{2-9}
& \multirow{3}{*}{LogN}        & 10 & 0.81 & 0.50 & 0.36    & 0.77 & 0.48 & 0.31
\\
&                             & 1  & 0.77 & 0.45 & 0.36    & 0.73 & 0.42 & 0.31
\\
&                             & 0.1  & 0.72 & 0.41 & 0.36    & 0.70 & 0.39 & 0.31
\\\hline\hline
\multirow{6}{*}{Factors} &
\multirow{3}{*}{Normal} & 10 & 0.90 & 0.90 & 0.74    & 0.90 & 0.90 & 0.72
\\
&                             & 1  & 0.82 & 0.83 & 0.74    & 0.81 & 0.82 & 0.72
\\
&                             & 0.1  & 0.75 & 0.76 & 0.74    & 0.74 & 0.74 & 0.72
\\\cline{2-9}
&\multirow{3}{*}{LogN}        & 10 & 0.83 & 0.54 & 0.31    & 0.81 & 0.57 & 0.26
\\
&                             & 1  & 0.80 & 0.53 & 0.31    & 0.77 & 0.50 & 0.26
\\
&                             & 0.1  & 0.75 & 0.48 & 0.31    & 0.74 & 0.46 & 0.26
\\\hline

\end{tabular}
\end{center}


\end{table}

\begin{table}[htbp]
\footnotesize
\begin{center}
\caption{{\bf In-sample Estimation}: Median of 5 canonical correlations of the estimated loadings/factors and the true
ones when $N=60,T=150$: the larger the better }
\label{Supp_tab3}
\begin{tabular}{c|c|c|cccc|cccc}
\hline\hline
& &  &  \multicolumn{4}{c|}{Model (I)}
&  \multicolumn{4}{c}{Model (II)}
\\
&   $\bu_t$ &  $\omega$
  & SPCA & SPCA-LS & PCA & LS & SPCA & SPCA-LS & PCA &LS
\\\hline
\multirow{6}{*}{Loading}&
\multirow{3}{*}{Normal} & 10  & 0.95 & 0.95 & 0.88 & 0.82 & 0.93 & 0.93 & 0.85 & 0.78
\\
&                       & 1   & 0.92 & 0.92 & 0.88 & 0.83 & 0.88 & 0.88 & 0.85 & 0.79
\\
&                       & 0.1 & 0.85 & 0.86 & 0.88 & 0.86 & 0.84 & 0.84 & 0.85 & 0.83
\\\cline{2-11}
& \multirow{3}{*}{LogN} & 10  & 0.86 & 0.59 & 0.44 & 0.38 & 0.84 & 0.55 & 0.41 & 0.34
\\
&                       & 1   & 0.83 & 0.55 & 0.44 & 0.40 & 0.80 & 0.52 & 0.41 & 0.36
\\
&                       & 0.1 & 0.79 & 0.48 & 0.44 & 0.43 & 0.75 & 0.44 & 0.41 & 0.39
\\\hline\hline
\multirow{6}{*}{Factors}&
\multirow{3}{*}{Normal} & 10  & 0.94 & 0.94 & 0.83 & 0.75 & 0.91 & 0.91 & 0.81 & 0.74
\\
&                       & 1   & 0.86 & 0.86 & 0.83 & 0.78 & 0.83 & 0.83 & 0.81 & 0.76
\\
&                       & 0.1 & 0.81 & 0.82 & 0.83 & 0.82 & 0.79 & 0.79 & 0.81 & 0.79
\\\cline{2-11}
&\multirow{3}{*}{LogN}  & 10  & 0.85 & 0.66 & 0.40 & 0.33 & 0.84 & 0.64 & 0.37 & 0.30
\\
&                       & 1   & 0.81 & 0.60 & 0.40 & 0.35 & 0.80 & 0.61 & 0.37 & 0.32
\\
&                       & 0.1 & 0.77 & 0.54 & 0.40 & 0.38 & 0.75 & 0.56 & 0.37 & 0.35
\\\hline

\end{tabular}
\end{center}
\end{table}

\begin{table}[htbp]
\footnotesize
\begin{center}
\caption{{\bf Dependent data}: Median of  canonical correlations of the estimated loadings/factors and the true
ones when $N=40,T=100$: the larger the better }
\label{Dtab1_2}
\begin{tabular}{c|c|c|ccc|ccc}
\hline\hline
& &  &  \multicolumn{3}{c|}{Model (I)}
&  \multicolumn{3}{c}{Model (II)}
\\
&   $\bu_t$ &  $\omega$
  & SPCA & SPCA-LS & PCA & SPCA & SPCA-LS & PCA
\\\hline
\multirow{6}{*}{Loadings} &
\multirow{3}{*}{Normal}       & 10 & 0.89 & 0.90 & 0.78    & 0.87 & 0.87 & 0.73
\\
&                             & 1  & 0.84 & 0.84 & 0.78    & 0.82 & 0.82 & 0.73
\\
 &                            & 0.1 & 0.80 & 0.81 & 0.78    & 0.76 & 0.77 & 0.73
\\\cline{2-9}
& \multirow{3}{*}{LogN}       & 10 & 0.75 & 0.47 & 0.25    & 0.73 & 0.45 & 0.22
\\
&                             & 1  & 0.69 & 0.41 & 0.25    & 0.69 & 0.39 & 0.22
\\
&                             & 0.1  & 0.64 & 0.38 & 0.25    & 0.62 & 0.35 & 0.22
\\\hline\hline
\multirow{6}{*}{Factors} &
\multirow{3}{*}{Normal}       & 10 & 0.88 & 0.89 & 0.71    & 0.88 & 0.88 & 0.68
\\
&                             & 1  & 0.81 & 0.82 & 0.71    & 0.80 & 0.81 & 0.68
\\
&                             & 0.1  & 0.73 & 0.74 & 0.71    & 0.72 & 0.72 & 0.68
\\\cline{2-9}
&\multirow{3}{*}{LogN}        & 10 & 0.80 & 0.59 & 0.24    & 0.78 & 0.55 & 0.19
\\
&                             & 1  & 0.74 & 0.51 & 0.24    & 0.72 & 0.49 & 0.19
\\
&                             & 0.1  & 0.70 & 0.45 & 0.24    & 0.69 & 0.40 & 0.19
\\\hline

\end{tabular}
\end{center}
\end{table}

\section{Conclusions}\label{sec:con}

We study factor models when the factors depend on  observed explanatory characteristics.
The proposed method incorporates the explanatory power of these observed covariates,  and  is robust to possibly heavy-tailed distributions.
We focus on the case $\dim(\bx_t)$ is finite, and on the rates of convergence for the estimated factors and loadings. Under various signal-noise ratios, substantial improved rates of convergence can be gained.

Related to the above, the idea could be easily extended to the case that $\dim(\bx_t)$ is slowly growing (with respect to $(N, T)$). On the other hand, allowing $\dim(\bx_t)$ to be fast-growing would require some dimension-reduction treatment combined with covariate selections.
In addition,  selecting the covariates would be also useful  as  the quality of the signal is crucial.
We shall leave these open questions for future studies.

\newpage
\appendix

\section{Proof of Theorem \ref{th0}}
\begin{proof}
 Let $\bxi_1,...,\bxi_N$ be the eigenvectors of $\bSigma_{y|x}$, corresponding to the eigenvalues
 $\lambda_1(\bSigma_{y|x})\geq\lambda_2(\bSigma_{y|x})...\geq\lambda_N(\bSigma_{y|x})$. Due to $\bSigma_{y|x}=\bLambda\bSigma_{f|x}\bLambda'$, and by the assumption that $\lambda_{\min}(\bSigma_{f|x})>0$,  the rank of $\bSigma_{y|x}$  equals $K.$ Hence $\lambda_i(\bSigma_{y|x})=0$ for all $i>K.$

 (i)
  Let
 $
\bL=\bSigma_{\Lambda,N}^{1/2} \bSigma_{f|x}\bSigma_{\Lambda,N}^{1/2}.
 $
 Let $\bM$ be a $K\times K$ matrix, whose columns are the eigenvectors of $\bL$. Then   $\bD:=\bM' \bL\bM$ is a diagonal matrix, with diagonal elements being the eigenvalues of   $\bL$.  Let
$
\bH=\bSigma_{\Lambda,N}^{-1/2}\bM.
$
Then
$$
\frac{1}{N}\bSigma_{y|x}\bLambda\bH=\bLambda\bSigma_{f|x}\bSigma_{\Lambda,N}\bH
= \bLambda \bSigma_{\Lambda,N}^{-1/2}\bL\bM
\underbrace{=}_{\bM\bM'=\bI}\bLambda\bH\bM' \bL\bM=\bLambda\bH\bD.
$$
In addition,  $(\bLambda\bH)'(\bLambda\bH)=N\bM'\bM=N\bI_K,
$
 hence the columns of $\bLambda\bH/\sqrt{N}$ are the eigenvectors of $ \bSigma_{y|x}$, corresponding to the  $K$ nonzero eigenvalues.

(ii) From $E(\by_t|\bx_t)=\bLambda \bg(\bx_t)$, we have $(\bLambda\bH)' E(\by_t|\bx_t)=(\bLambda\bH)'\bLambda\bH\bH^{-1} \bg(\bx_t)$. This leads to the desired expression of $\bH^{-1} \bg(\bx_t)$.

(iii) The nonzero eigenvalues  of $\bSigma_{y|x}=\bLambda\bSigma_{f|x}\bLambda'$  equal  those of
$$\bSigma_{f|x}^{1/2}\bLambda' \bLambda \bSigma_{f|x}^{1/2}=N\bSigma_{f|x}^{1/2}\bSigma_{\Lambda, N}\bSigma_{f|x}^{1/2},
$$
which are also the same as those of $N\bSigma_{\Lambda,N}^{1/2}\bSigma_{f|x}\bSigma_{\Lambda,N}^{1/2}=N\bL$.
Note that
$$
\lambda_{\min}(N\bL)\geq N\lambda_{\min}(\bSigma_{f|x}) \lambda_{\min}(\bSigma_{\Lambda, N})\geq N\chi_N\underbar{c}_\Lambda.
$$

 \end{proof}

\section{Proofs for Section 4}
\subsection{A bird's-eye view  of the major technical   steps}

We first provide a  bird's-eye view  of the major steps in the proof.
 The key intermediate result is to prove the following Bahadur representation of the estimated eigenvectors:
\begin{eqnarray}\label{eb.1}
\widehat\bLambda-\bLambda\bH &=&\frac{1}{NT}\sum_{t=1}^T\bLambda \bg(\bx_t)\Phi(\bx_t)'\bA\sum_{i=1}^N\frac{1}{T}\sum_{s=1}^T\Phi(\bx_s)'\dot{\rho}(\alpha_T^{-1}e_{is})\alpha_T \widehat\bLambda \widetilde\bV^{-1} \cr
&&  + \widetilde{\bDelta}(\{\bx_t, \be_t\}_{t\leq T}),
\end{eqnarray}
for some invertible matrix $\bH$.
Here the first term on the right hand side is the leading term that results in the presented rate of convergence in Theorem \ref{t31}, where $\dot{\rho}(\cdot)$ denotes the derivative of Huber's loss function; $\widetilde\bV$ is a $K$-dimensional diagonal matrix of the eigenvalues of $\widehat\bSigma/N$. The second term $ \widetilde{\bDelta}(\{\bx_t, \be_t\}_{t\leq T})$ is a higher order random term that depends on both $\{\bx_t\}$ and $\{ \be_{t}\}$, where   $\be_t=\by_t-E(\by_t|\bx_t)=(e_{1t},...,e_{Nt})$. 

To have an general idea of how we prove (\ref{eb.1}),
recall that $\widehat\bSigma/N:=\frac{1}{TN}\sum_{t=1}^T\widehat E(\by_t|\bx_t)\widehat E(\by_t|\bx_t)'$, where  each element of  $\widehat E(\by_t|\bx_t)$ is  $\widehat E(y_{it}|\bx_t)=\widehat\bb_i'\Phi(\bx_t)$ with $\widehat\bb_i$ being the M-estimator of the sieve coefficients of $E(y_{it}|\bx_t)$, obtained by minimizing the Huber's loss:$$
\widehat\bb_i=\arg\min_{\bb\in\mathbb{R}^J}Q_i(\bb),\qquad
Q_i(\bb)=\frac{1}{T}\sum_{t=1}^T\alpha_T^2\rho\left(\frac{y_{it}-\Phi(\bx_t)'\bb}{\alpha_T}\right).
$$
 Then
   by the definition of $\widehat\bLambda$,
\begin{equation}\label{eb.2new}
\frac{1}{N}\widehat\bSigma \widehat\bLambda=\widehat\bLambda\widetilde\bV.
\end{equation}
The above is the key equality we shall use to derive (\ref{eb.1}).
To   use this equality, we need to obtain the Bahadur representations of $\widehat\bb_i$ and $\widehat E(y_{it}|\bx_t)$ in the following steps.

\textbf{Step 1: bias of sieve coefficients. } Define, for $i=1,...,N$,
$$\bb_i:=\arg\min_{\bb\in\mathbb{R}^J} E[y_{it}-\bb'\Phi(\bx_t)]^2, \quad
\bb_{i,\alpha}=\arg\min_{\bb\in\mathbb{R}^J} E\alpha_T^2\rho\left( \frac{y_{it}-\Phi(\bx_t)'\bb}{\alpha_T}\right).
$$ Note that the sieve expansion of $ E(\by_t|\bx_t)$ is $ \bb_i'\Phi(\bx_t)$ (to be proved in Lemma \ref{lb.1ad}). But
 $\widehat\bb_i$ is biased for estimating $\bb_i$, and asymptotically converges to $\bb_{i,\alpha}$.
  As $\alpha_T\to\infty$, $\bb_{i,\alpha}$ is expected to converge to $\bb_i$ uniformly in $i\leq N$. This is true given some moment conditions on
$
\be_t:=\by_t-E(\by_t|\bx_t).
$

\textbf{Step 2: Expansion of $\widehat\bb_i-\bb_{i,\alpha}$.}

 The first order condition gives $\nabla Q_i(\widehat\bb_i)=0.$  But  we cannot directly expand this equation because  $\nabla Q_i$ is not differentiable.  As in  many M-estimations,  define $\bar Q_i(\bb)=EQ_i(\bb)$, and $\bmu_i(\bb)=\nabla Q_i(\bb)-\nabla \bar Q_i(\bb)$.   So we have
 $$
0= \nabla \bar Q_i(\widehat\bb_i) - \bmu_i(\widehat \bb_i),
 $$
and $\nabla \bar Q_i$ is differentiable.   We shall apply  the standard empirical process theory for independent data
(the symmetrization  and contraction  theorems, e.g.,    \cite{bulmann} )   to prove the stochastic equicontinuity of $\bmu_i(\bb)$ and thus the convergence of   $\max_i\|\bmu_i(\widehat\bb_i)-\bmu_i(\bb_{i,\alpha})\|  $. This will eventually lead to an expansion  of $\widehat\bb_i-\bb_{i,\alpha}$,  to be given in Lemma \ref{lb1}.

 \textbf{Step 3: Expansion of $\widehat E(\by_t|\bx_t) - E(\by_t|\bx_t)$.}
Combining steps 1 and 2 will eventually lead to
\begin{equation}\label{eb.3new}
\widehat E(y_{it}|\bx_t)=E(y_{it}|\bx_t)+ \Phi(\bx_t)'\bA\frac{1}{T}\sum_{s=1}^T\alpha_T\dot{\rho}(\alpha_T^{-1}e_{is})\Phi(\bx_s)+ R_{it}
\end{equation}
where $R_{it}$ is a high-order remainder term that depends on $\bx_t$, and $\bA=(2E\Phi(\bx_t)\Phi(\bx_t)')^{-1}$ is the Hessian matrix. We shall bound $\max_{i\leq N}\frac{1}{T}\sum_{t=1}^TR_{it}$  in Proposition \ref{pb3}.

  \textbf{Step 4: Expansion of $\widehat\bLambda-\bLambda\bH $.}
  Substituting the expansion of $\widehat E(y_{it}|\bx_t)  $ to (\ref{eb.3new}), with $\widehat E(\by_t|\bx_t)$ replaced by its expansions, we will eventually obtain (\ref{eb.1}). Then  (\ref{eb.1}) can be directly applied to obtain the rate of convergence for the estimated loadings.    This will be done in Section \ref{sd.3}, where  we show that the remainder term is of  a smaller order than the leading term.

  Importantly, both the signal strength  $\chi_N=\lambda_{\min}(\bSigma_{f|x})$ and the ``noise" $\cov(\bgamma_t)$ plays an essential role in (\ref{eb.2new}), which are to be reflected in the rate of convergence.

\subsection{Estimating the loadings} 

Throughout the proofs, as $T, J\to\infty$,  $N$ either grows or stays constant.

Write $\bM_{\alpha}$ be an $N\times J$ matrix, whose $i$th row is given by
$$
\bM_{i,\alpha}':=\frac{1}{T}\sum_{s=1}^T\alpha_T\dot{\rho}(\alpha_T^{-1}e_{is})\Phi(\bx_s)'.
$$
Write $\bR_t=(R_{1t},...,R_{Nt})'$, where $R_{it}$ was defined in Proposition \ref{pb3}.
Then the Bahadur representation in Proposition \ref{pb3} can be written in the vector form:  $\bA=(2E\Phi(\bx_t)\Phi(\bx_t)')^{-1}$,
\begin{equation}\label{eb1}
\widehat E(\by_t|\bx_t)=E(\by_t|\bx_t)+\bM_{\alpha}\bA\Phi(\bx_t)+\bR_t=\bLambda E(\bff_t|\bx_t)+\bM_{\alpha}\bA\Phi(\bx_t)+\bR_t.
\end{equation}
Let $\widetilde\bV$ be a $K\times K$ diagonal matrix, whose diagonal elements are the first $K$ eigenvalues of $\widehat\bSigma/N:=\frac{1}{TN}\sum_{t=1}^T\widehat E(\by_t|\bx_t)\widehat E(\by_t|\bx_t)'$.
 By the definition of $\widehat\bLambda$,
$
\frac{1}{N}\widehat\bSigma \widehat\bLambda=\widehat\bLambda\widetilde\bV.
$
Plugging in (\ref{eb1}), with $\widehat\bSigma=\frac{1}{T}\sum_{t=1}^T\widehat E(\by_t|\bx_t)\widehat E(\by_t|\bx_t)'$
we have,
\begin{equation}\label{eb.2add}
\widehat\bLambda -\bLambda \bH=\sum_{i=1}^8 \bB_i,\qquad \bH=\frac{1}{TN}\sum_{t=1}^TE(\bff_t|\bx_t)E(\bff_t|\bx_t)'\bLambda'\widehat\bLambda\widetilde\bV^{-1}
\end{equation}
where  for $\bA=(2E\Phi(\bx_t)\Phi(\bx_t)')^{-1}$,
 \begin{eqnarray*}
\bB_1&=&\frac{1}{TN}\sum_{t=1}^T 	\bLambda E(\bff_t|\bx_t)\Phi(\bx_t)'\bA\bM_{\alpha}'\widehat\bLambda		 \widetilde\bV^{-1} ,\qquad \bB_2=\frac{1}{TN}\sum_{t=1}^T \bLambda E(\bff_t|\bx_t)\bR_t'\widehat\bLambda			 \widetilde\bV^{-1},\cr
 \bB_3&=&\frac{1}{TN}\sum_{t=1}^T \bM_{\alpha}\bA\Phi(\bx_t)	E(\bff_t|\bx_t)'\bLambda'	\widehat\bLambda	 \widetilde\bV^{-1} \qquad \bB_4=\frac{1}{TN}\sum_{t=1}^T \bM_{\alpha}\bA\Phi(\bx_t)\Phi(\bx_t)'\bA\bM_{\alpha}'\widehat\bLambda			 \widetilde\bV^{-1},\cr
  \bB_5&=&\frac{1}{TN}\sum_{t=1}^T 	\bM_{\alpha}\bA\Phi(\bx_t)\bR_t'\widehat\bLambda		 \widetilde\bV^{-1},\qquad \bB_6=\frac{1}{TN}\sum_{t=1}^T 	\bR_tE(\bff_t|\bx_t)'\bLambda'\widehat\bLambda		 \widetilde\bV^{-1} ,\cr
   \bB_7&=&\frac{1}{TN}\sum_{t=1}^T 	\bR_t\Phi(\bx_t)'\bA\bM_{\alpha}'\widehat\bLambda		 \widetilde\bV^{-1} ,\qquad \bB_8=\frac{1}{TN}\sum_{t=1}^T 	\bR_t\bR_t'\widehat\bLambda		 \widetilde\bV^{-1} .
\end{eqnarray*}

We derive the rates of convergence by examining each term of (\ref{eb.2add}). 

\subsubsection{Proof of Theorem \ref{t31}: $\frac{1}{N}\sum_{i=1}^N\|\widehat\blambda_i-\bH'\blambda_i\|^2$}

\begin{prop} \label{pb4} Suppose $J^{3}\log^2N=O(T )$, $\eta\geq 2, $ and $J^2/T+J^{-\eta}\ll\chi_N$. Then
  $$
\frac{1}{N}\|\widehat\bLambda-\bLambda\bH\|_F^2=O_P(\frac{J}{T}+J^{1-2\eta})\chi_N^{-1}.$$

\end{prop}
\proof
From Lemma \ref{lb2} and Proposition \ref{pb3}, we obtain
\begin{eqnarray*}
 \frac{1}{N}\|\bM_{\alpha}\|^2+ \max_i\frac{1}{T}\sum_{t=1}^TR_{it}^2&=&O_P(\frac{J}{T}+J^{1-2\eta}+ \alpha_T^{-2(\zeta_1-1)}\frac{J^3\log N}{T}+\frac{J^3\log N\log J}{T^2})\cr
 &\leq &O_P(\frac{J}{T}+J^{1-2\eta})
\end{eqnarray*}
under the assumption $J^{3 }\log^2N=O(T )$,  $\alpha_T=C\sqrt{T/\log (NJ)}$ and $\zeta_1>2$.
Hence from Lemma \ref{lb2} and Proposition \ref{pb3},
\begin{eqnarray*}
 \frac{1}{N}\|\widehat\bLambda-\bLambda\bH\|_F^2&=&O_P(\frac{1}{N}\sum_{i=1}^8\|\bB_i\|_F^2)\cr
 &=&O_P(\frac{1}{N}\|\bM_{\alpha}\|^2J\max_i\frac{1}{T}\sum_{t=1}^TR_{it}^2/\chi_N^2)\cr
&&+O_P(\frac{1}{N}\|\bM_{\alpha}\|^2\chi_N^{-1}+\frac{1}{N}\|\bM_{\alpha}\|_F^4/(N\chi_N^2))\cr
&&+O_P( \max_i\frac{1}{T}\sum_{t=1}^TR_{it}^2   \chi_N^{-1}+ (\max_i\frac{1}{T}\sum_{t=1}^TR_{it}^2)^2/\chi_N^2)\cr
&\leq& O_P( \chi_N^{-2} J) ( \frac{1}{N}\|\bM_{\alpha}\|^2+ \max_i\frac{1}{T}\sum_{t=1}^TR_{it}^2 )^2
\cr
&&+O_P( \chi_N^{-1}  ) ( \frac{1}{N}\|\bM_{\alpha}\|^2 +\max_i\frac{1}{T}\sum_{t=1}^TR_{it}^2 ) \cr
&\leq & O_P(\frac{J}{T}+J^{1-2\eta})\chi_N^{-1}[1+(\frac{J}{T}+J^{1-2\eta})\chi_N^{-1}J]\cr
&\leq &O_P(\frac{J}{T}+J^{1-2\eta})\chi_N^{-1}.
\end{eqnarray*}
The last equality is due to $(\frac{J}{T}+J^{1-2\eta})\chi_N^{-1}J=O(1)$, granted by $\eta\geq 2, $ and $J^2/T+J^{-\eta}\ll\chi_N$.
Q.E.D.

 \subsubsection{Proof of Theorem \ref{t31}: $\max_{i\leq N}\|\blambda_i-\bH'\blambda_i\|$}

 \begin{eqnarray*}
\bB_1&=&\frac{1}{TN}\sum_{t=1}^T 	\bLambda E(\bff_t|\bx_t)\Phi(\bx_t)'\bA\bM_{\alpha}'\widehat\bLambda		 \widetilde\bV^{-1} ,\qquad \bB_2=\frac{1}{TN}\sum_{t=1}^T \bLambda E(\bff_t|\bx_t)\bR_t'\widehat\bLambda			 \widetilde\bV^{-1},\cr
 \bB_3&=&\frac{1}{TN}\sum_{t=1}^T \bM_{\alpha}\bA\Phi(\bx_t)	E(\bff_t|\bx_t)'\bLambda'	\widehat\bLambda	 \widetilde\bV^{-1} \qquad \bB_4=\frac{1}{TN}\sum_{t=1}^T \bM_{\alpha}\bA\Phi(\bx_t)\Phi(\bx_t)'\bA\bM_{\alpha}'\widehat\bLambda			 \widetilde\bV^{-1},\cr
  \bB_5&=&\frac{1}{TN}\sum_{t=1}^T 	\bM_{\alpha}\bA\Phi(\bx_t)\bR_t'\widehat\bLambda		 \widetilde\bV^{-1},\qquad \bB_6=\frac{1}{TN}\sum_{t=1}^T 	\bR_tE(\bff_t|\bx_t)'\bLambda'\widehat\bLambda		 \widetilde\bV^{-1} ,\cr
   \bB_7&=&\frac{1}{TN}\sum_{t=1}^T 	\bR_t\Phi(\bx_t)'\bA\bM_{\alpha}'\widehat\bLambda		 \widetilde\bV^{-1} ,\qquad \bB_8=\frac{1}{TN}\sum_{t=1}^T 	\bR_t\bR_t'\widehat\bLambda		 \widetilde\bV^{-1} .
\end{eqnarray*}

 \proof By Lemma \ref{lb4}
$\max_{i\leq N}\|\bM_{i,\alpha}\|=O_P(J^{-\eta}\sqrt{J}+\sqrt{J(\log N)/T})$.
 Let $\bB_{i1},...,\bB_{i8}$ respectively denote the $i$th row of $\bB_1,...,\bB_8$. 
We have
\begin{eqnarray*}
 \max_i\|\bB_{i1}\|&\leq& \chi_N^{-1/2}   O_P( \|\bM_{\alpha}\widehat\bLambda\|/N)\leq O_P(\chi_N^{-1/2}\max_i\|\bM_{i,\alpha}\|)\cr
  \max_i\|\bB_{i2}\|&\leq& \chi_N^{-1/2}O_P(\max_i\frac{1}{T}\sum_{t=1}^TR_{it}^2)^{1/2})\cr
  \max_i\|\bB_{i3}\|&\leq&\chi_N^{-1/2}O_P(\max_i\|\bM_{i,\alpha}\|)=O_P(J^{-\eta}\sqrt{J}+\sqrt{J(\log N)/T})\chi_N^{-1/2}\cr
    \max_i\|\bB_{i4}\|&\leq&O_P( \max_i\|\bM_{i,\alpha}\|)O_P(\|\bM_{\alpha}'\widehat\bLambda\|/N)\chi_N^{-1}\cr
       \max_i\|\bB_{i5}\|&\leq&O_P(\max_i\|\bM_{i,\alpha}\|)O_P(\sqrt{J}\max_i\frac{1}{T}\sum_tR_{it}^2)^{1/2}\chi_N^{-1}  \cr
              \max_i\|\bB_{i6}\|&\leq&O_P(\max_i\frac{1}{T}\sum_tR_{it}^2)^{1/2})\chi_N^{-1/2}\cr
       \max_i\|\bB_{i7}\|&\leq&     O_P(\max_i\frac{1}{T}\sum_tR_{it}^2)^{1/2}\sqrt{J})O_P( \|\bM_{\alpha}\widehat\bLambda\|/N)\chi_N^{-1}\cr
          \max_i\|\bB_{i8}\|&\leq& O_P(   \max_i\frac{1}{T}\sum_tR_{it}^2 \chi_N^{-1}  ).
\end{eqnarray*}
 Hence
\begin{eqnarray*}
&&\max_{i\leq N}\|\blambda_i-\bH'\blambda_i\|\leq O_P(   \max_i\|\bB_{i2}\| + \max_i\|\bB_{i3}\|  ) \cr
&=&O_P(J^{-\eta}\sqrt{J}+\sqrt{J(\log N)/T}+\alpha_T^{-(\zeta_1-1)} \sqrt{\frac{J^3\log N}{T}}) \chi_N^{-1/2} \cr
&=&O_P(J^{-\eta}\sqrt{J}+\sqrt{J(\log N)/T}) \chi_N^{-1/2} ,
\end{eqnarray*}
where the last equality follows from
$$\alpha_T^{-(\zeta_1-1)} \sqrt{\frac{J^3\log N}{T}}=O(\sqrt{\frac{J\log N}{T}})$$
under assumptions $(\log N)^2 J^3=O(T)$ and $\zeta_1>2$.

  \subsection{Proof of Theorem \ref{t32}:  factors}\label{sd.4}

Recall that $ \widehat\bg(\bx_t)=\frac{1}{N}\widehat\bLambda' \widehat E(\by_t|\bx_t) $. By (\ref{eb1}),   $ \widehat\bg(\bx_t)- \bH^{-1}\bg(\bx_t)=\sum_{i=1}^4\bC_{ti}$, where
 \begin{eqnarray*}
 \bC_{t1}&=& \frac{1}{N}(\widehat\bLambda-\bLambda\bH)' (\bLambda\bH -\widehat\bLambda)\bH^{-1}E(\bff_t|\bx_t),\qquad   \bC_{t3}= \frac{1}{N}\widehat\bLambda '\bM_{\alpha}\bA\Phi(\bx_t), \cr
\bC_{t2}&=& -\frac{1}{N} \bH'\bLambda' ( \widehat\bLambda-\bLambda\bH)\bH^{-1}E(\bff_t|\bx_t)      , \qquad \bC_{t4}=\frac{1}{N}\widehat\bLambda'\bR_t.
 \end{eqnarray*}



The convergence of  $ \frac{1}{T}\sum_{t=1}^T\|\widehat\bg(\bx_t)-\bH^{-1}\bg(\bx_t)\|^2$  in this theorem is proved in the following proposition.

 \begin{prop}\label{pb6} As $T\to\infty$ and $N$  either grows or stays constant,
\begin{eqnarray*}
&& \frac{1}{T}\sum_{t=1}^T\|\widehat\bg(\bx_t)-\bH^{-1}\bg(\bx_t)\|^2= O_P(  \frac{J^2}{T^2}\chi_N^{-1}   +\frac{J\| \cov(\bgamma_s)\|}{T}+J^{1-2\eta}+\frac{J}{TN}+\frac{J^3\log^2 N}{T^2}) .
 \end{eqnarray*}

\end{prop}

 \proof  Recall
$$
a_{T}^2:=\frac{J}{T}+J^{1-2\eta},\quad b_{NT}^2:=\frac{J\| \cov(\bgamma_s)\|}{T}+\frac{J}{TN}+\frac{ J}{T}\alpha_T^{-\zeta_2}.
$$
  By Lemma \ref{lb.8}, $\|\bH\|=O_P(1)=\|\bH^{-1}\|$. Also, by
 Proposition \ref{pb4} and Lemmas \ref{lb3}, \ref{lb5},
  \begin{eqnarray*}
&&\frac{1}{N}\|\widehat\bLambda-\bLambda\bH\|_F^2= O_P(a_T^2\chi_N^{-1})  . \cr
&&\frac{1}{N}\|\bM_{\alpha}'\widehat\bLambda\|_F=    O_P(a_T^2\chi_N^{-1/2}) +O_P(b_{NT})   \cr
&& \|\frac{1}{N}\bLambda'(\widehat\bLambda-\bLambda\bH)\|\leq    O_P(\chi_N^{-1/2})  ( \frac{1}{N}\|\bM_{\alpha}'\widehat\bLambda\|_F+ (\max_i\frac{1}{T}\sum_{t=1}^TR_{it}^2)^{1/2} )
\end{eqnarray*}

Therefore,  as $\frac{1}{T}\sum_t\|E(\bff_t|\bx_t)\|^2=O_P(\chi_N)$,
 \begin{eqnarray*}
 \frac{1}{T}\sum_{t=1}^T\|\bC_{t1}\|^2&\leq &O_P(1)
[ \frac{1}{N}\|\widehat\bLambda-\bLambda\bH\|^2]^2 \chi_N
 \leq  O_P(a_T^4   \chi_N^{-1})
\cr
 \frac{1}{T}\sum_{t=1}^T\|\bC_{t2}\|^2&\leq& O_P(1)[\frac{1}{N} \bLambda' ( \widehat\bLambda-\bLambda\bH)  ]^2\chi_N
\cr
& \leq& O_P  (a_T^4\chi_N^{-1}+b_{NT}^2+ \max_i\frac{1}{T}\sum_{t=1}^TR_{it}^2  )
 \cr
    \frac{1}{T}\sum_{t=1}^T\|\bC_{t4}\|^2&=&O_P(\max_i\frac{1}{T}\sum_{t=1}^TR_{it}^2).
    \end{eqnarray*}
        Finally, let $\bbeta_i$ denote the $i$th row of $\frac{1}{N}\widehat\bLambda'\bM_{\alpha}\bA$, $i\leq K$. Then
\begin{eqnarray*} && \frac{1}{T}\sum_{t=1}^T\|\bC_{t3}\|^2 = \frac{1}{T}\sum_{t=1}^T\|\frac{1}{N}\widehat\bLambda'\bM_{\alpha}\bA\Phi(\bx_t)\|^2= \sum_{i=1}^K\frac{1}{T}\sum_{t=1}^T(\bbeta_i'\Phi(\bx_t))^2\cr
&\leq& \sum_{i=1}^K\|\bbeta_i\|^2 \| \frac{1}{T}\sum_{t=1}^T\Phi(\bx_t)\Phi(\bx_t)'\| \cr
&=& O_P(1)\|\frac{1}{N}\widehat\bLambda'\bM_{\alpha}\bA\|_F^2=O_P(\frac{1}{N^2}\|\widehat\bLambda'\bM_{\alpha}\|^2)\cr
&\leq&O_P(b_{NT}^2 + a_T^4\chi_N^{-1}).
    \end{eqnarray*}
Thus
    \begin{eqnarray*}
&& \frac{1}{T}\sum_{t=1}^T\|\widehat\bg(\bx_t)-\bH^{-1}\bg(\bx_t)\|^2\leq O_P(1)\sum_{i=1}^4\frac{1}{T}\sum_{t=1}^T\|\bC_{ti}\|^2
\cr
&\leq&  O_P  (a_T^4\chi_N^{-1}+b_{NT}^2+ \max_i\frac{1}{T}\sum_{t=1}^TR_{it}^2  ) \cr
&\leq& O_P(  \frac{J^2}{T^2}\chi_N^{-1}+J^{2-4\eta} \chi_N^{-1}  +J^{1-2\eta}+\frac{J\| \cov(\bgamma_s)\|}{T}+\frac{J}{TN}+\frac{J^3\log N\log J}{T^2}\cr
&&+\frac{ J}{T}\alpha_T^{-\zeta_2}+ \alpha_T^{-2(\zeta_1-1)}\frac{J^3\log N}{T}  )\cr
&\leq^{(1)}&  O_P(  \frac{J^2}{T^2}\chi_N^{-1}   +\frac{J\| \cov(\bgamma_s)\|}{T}+J^{1-2\eta}+\frac{J}{TN}+\frac{J^3\log^2 N}{T^2})  
 \end{eqnarray*}
where (1) is due to $\zeta_1, \zeta_2>2$, and  $J^{3}\log^2N=O(T )$,
 $$
\frac{ J}{T}\alpha_T^{-\zeta_2}+ \alpha_T^{-2(\zeta_1-1)}\frac{J^3\log N}{T}  + \frac{J^3\log N\log J}{T^2}=O(\frac{J^3\log^2 N }{T^2})
 $$
and $\chi_N\gg J^{-\eta}$ (so $J^{2-4\eta}\chi_N^{-1}=O(J^{1-2\eta})$). Q.E.D.

 
\begin{prop}
$$
 \frac{1}{T}\sum_{t=1}^T\|    \widehat\bgamma_t-\bH^{-1}\bgamma_t
\|^2=   O_P(\frac{1}{N})  + O_P(\chi_N^{-1})(   \frac{ J^4(\log N)^2}{T^2} +\frac{J^2}{T^2}\chi_N^{-1}   +\frac{J\| \cov(\bgamma_s)\|}{T}+J^{1-2\eta}+\frac{J}{TN}).
$$
\end{prop}

 Note that $
 \by_t-E(\by_t|\bx_t)=\bLambda\bgamma_t+\bu_t.$
and
$
 \widehat\bgamma_t=\frac{1}{N} \widehat\bLambda'(\by_t-\widehat E(\by_t|\bx_t)). $ Hence from (\ref{eb1})
\begin{equation}\label{eb.2}
   \widehat\bgamma_t-\bH^{-1}\bgamma_t=\frac{1}{N}\bH'\bLambda'\bu_t+  \bD_{t1} +\bD_{t2}+\bC_{t3} +\bC_{t4}
\end{equation}
    where  $\bC_{t3}, \bC_{t4}$ are as  defined earlier, and
    \begin{eqnarray*}
    \bD_{t1}&=&  \frac{1}{N} \widehat\bLambda'(\bLambda\bH-\widehat\bLambda)\bH^{-1}\bgamma_t,\qquad 	 \bD_{t2}= \frac{1}{N} (\widehat\bLambda-\bLambda\bH)'\bu_t \cr
      \bC_{t3}&=&\frac{1}{N}\widehat\bLambda'\bM_{\alpha}\bA\Phi(\bx_t) ,\qquad   \bC_{t4}=\frac{1}{N}\widehat\bLambda'\bR_t  .
        \end{eqnarray*}
Hence  for a constant $C>0$,
 $ \frac{1}{T}\sum_{t=1}^T\|    \widehat\bgamma_t-\bH^{-1}\bgamma_t
\|^2\leq C(\sum_{i=1}^2\frac{1}{T}\sum_{t=1}^T\|\bD_{ti}\|^2+\sum_{i=3}^4\frac{1}{T}\sum_{t=1}^T\|\bC_{ti}\|^2).
$
        We look at   terms  on the right hand side one by one. First of all,
        \begin{eqnarray*}
  E      \|  \frac{1}{T}\sum_{t=1}^T\bgamma_t\bgamma_t'-\cov(\bgamma_t)\|^2 _F&=&\sum_{i=1}^K\sum_{j=1}^K\var(\frac{1}{T}\sum_{t=1}^T\gamma_{it}\gamma_{jt} )\cr
&=&\sum_{i=1}^K\sum_{j=1}^K\frac{1}{T}\var(\gamma_{it}\gamma_{jt})\cr
&=&O(T^{-1})\max_{i, j\leq K}\var(\gamma_{it}\gamma_{jt}).
        \end{eqnarray*}
This implies
$
\| \frac{1}{T}\sum_{t=1}^T\bgamma_t\bgamma_t'\|\leq O_P(c_T)
$
where $$c_T:=\|\cov(\bgamma_t)\|+ (\frac{1}{T}\max_{i, j\leq K}\var(\gamma_{it}\gamma_{jt}))^{1/2}.$$

        As for $\bD_{t1}$, let $\bG= \frac{1}{N} \widehat\bLambda'(\bLambda\bH-\widehat\bLambda)\bH^{-1}$ and let $\bG_i'$ denote its $i$th row, $i\leq K$.  By (\ref{eb.5}),  and $\|\bH^{-1}\|=O_P(1)$,
        \begin{eqnarray*}
\|\bG\| ^2
&\leq& O_P(\chi_N^{-1})  ( \frac{1}{N}\|\bM_{\alpha}'\widehat\bLambda\|_F+ (\max_i\frac{1}{T}\sum_{t=1}^TR_{it}^2)^{1/2} )  ^2+O_P(a_T^4\chi_N^{-2}).
  \end{eqnarray*}
   Then
    \begin{eqnarray*}
\frac{1}{T}\sum_{t=1}^T\|\bD_{t1}\|^2&=&   \sum_{i=1}^K \frac{1}{T}\sum_{t=1}^T(\bG_i'\bgamma_t)^2=   \sum_{i=1}^K \bG_i'\frac{1}{T}\sum_{t=1}^T\bgamma_t\bgamma_t'\bG_i\cr
&\leq& \|\frac{1}{T}\sum_{t=1}^T\bgamma_t\bgamma_t'\|\|\bG\|_F^2\cr
&=&\|\bG\|_F^2O_P(c_T)\cr
  &\leq&
  c_TO_P(\chi_N^{-1})  ( \frac{1}{N}\|\bM_{\alpha}'\widehat\bLambda\|_F+ (\max_i\frac{1}{T}\sum_{t=1}^TR_{it}^2)^{1/2} )  ^2+O_P(c_Ta_T^4\chi_N^{-2}).
            \end{eqnarray*}

Terms $\bC_{t3}$ and $\bC_{t4}$ were bounded in the proof of Proposition \ref{pb6}:
    \begin{eqnarray*}
&&\frac{1}{T}\sum_{t=1}^T\|\bC_{t3}\|^2  +   \frac{1}{T}\sum_{t=1}^T\|\bC_{t4}\|^2   \leq O_P(\max_i\frac{1}{T}\sum_{t=1}^TR_{it}^2+\frac{1}{N^2}\|\widehat\bLambda'\bM_{\alpha}\|^2)     .
       \end{eqnarray*}

 Term $\bD_{t2}$ is given in Lemma \ref{lb8} below:
 \begin{eqnarray*}
  &&\frac{1}{T}\sum_{t=1}^T\|\bD_{t2}\|^2\cr
  & =&
  O_P(\chi_N^{-1})(\frac{1}{N^3}\|\bM_{\alpha}'\widehat\bLambda	  \|^2+\frac{1}{N}   \max_i\frac{1}{T}\sum_{t=1}^T   R_{it}^2 + \frac{1}{N^2T} \sum_{s=1}^T\|\bu_s'	 \bM_{\alpha}\|^2  +\frac{1}{N^2T^2} \sum_{s=1}^T         \sum_{t=1}^T| \bu_s'	\bR_t|^2)  .
  \end{eqnarray*}

By  Lemmas  \ref{lb6}, \ref{lb7},
 \begin{eqnarray*}
\frac{1}{N^2T} \sum_{s=1}^T\|\bu_s'	\bM_{\alpha}\|^2&\leq&O_P(\frac{J\|\cov(\bgamma_s)\|}{TN}+\frac{J}{T^2}+\frac{J}{N^2T}+\frac{J }{T}\alpha_T^{-\zeta_2})  \cr
\frac{1}{N^2T^2} \sum_{s=1}^T	 \sum_{t=1}^T 	|\bu_s'\bR_t|^2&\leq&
 O_P(\frac{ J^4\log N\log J}{T^2}+ \frac{J^{2-2\eta}   }{N}+\frac{ J^4\log N}{T}\alpha_T^{-2(\zeta_1-1)} )  .
  \end{eqnarray*}

  So combined with  Lemmas \ref{lb3}, Proposition \ref{pb3},
    \begin{eqnarray*}
&&\sum_{i=3}^4\frac{1}{T}\sum_{t=1}^T\|\bC_{ti}\|^2+\sum_{i=1}^2\frac{1}{T}\sum_{t=1}^T\|\bD_{ti}\|^2 =   O_P(c_Ta_T^4\chi_N^{-2})\cr
 &&+ O_P(1+c_T\chi_N^{-1}+N^{-1}\chi_N^{-1})(\frac{1}{N^2}\|\bM_{\alpha}'\widehat\bLambda	  \|^2+    \max_i\frac{1}{T}\sum_{t=1}^T   R_{it}^2 ) \cr
 &&+O_P(\chi_N^{-1})(  \frac{1}{N^2T} \sum_{s=1}^T\|\bu_s'	 \bM_{\alpha}\|^2  +\frac{1}{N^2T^2} \sum_{s=1}^T         \sum_{t=1}^T| \bu_s'	\bR_t|^2)  \cr
 &\leq^{(1)}&
 O_P(\chi_N^{-1})(\frac{1}{N^2}\|\bM_{\alpha}'\widehat\bLambda	  \|^2+    \max_i\frac{1}{T}\sum_{t=1}^T   R_{it}^2 +  \frac{1}{N^2T} \sum_{s=1}^T\|\bu_s'	 \bM_{\alpha}\|^2  +\frac{1}{N^2T^2} \sum_{s=1}^T         \sum_{t=1}^T| \bu_s'	\bR_t|^2) \cr
 &&+O_P(c_Ta_T^4\chi_N^{-2})\cr
 &\leq^{(2)}& O_P(\chi_N^{-1})(   \frac{ J^4(\log N)^2}{T^2} +\frac{J^2}{T^2}\chi_N^{-1}   +\frac{J\| \cov(\bgamma_s)\|}{T}+J^{1-2\eta}+\frac{J}{TN}
 +a_T^4\chi_N^{-1})
 \cr
  &\leq^{(3)}& O_P(\chi_N^{-1})(   \frac{ J^4(\log N)^2}{T^2} +\frac{J^2}{T^2}\chi_N^{-1}   +\frac{J\| \cov(\bgamma_s)\|}{T}+J^{1-2\eta}+\frac{J}{TN}
 ).
             \end{eqnarray*}
where (1) follows from that $1+c_T\chi_N^{-1}+\chi_N^{-1}N^{-1}=O(\chi_N^{-1})$  ; (2) is due to
 $
\frac{ J}{T}\alpha_T^{-\zeta_2}+ \alpha_T^{-2(\zeta_1-1)}\frac{J^4\log N}{T}  + \frac{J^4\log N\log J}{T^2}=O(\frac{J^4\log^2 N }{T^2})$ and that $c_T=O(1)$ due to Assumption \ref{a31}; (3) is due to $J^{-\eta}\chi_N^{-1}=O(1)$.

 Finally,
$
   \frac{1}{T}\sum_{t=1}^T\|  \frac{1}{N}\bH'\bLambda'\bu_t\|^2=O_P(\frac{1}{TN^2}\sum_{t=1}^TE\|  \bLambda'\bu_t\|^2)=O_P(\frac{1}{N}).
$
 Hence
$$
 \frac{1}{T}\sum_{t=1}^T\|    \widehat\bgamma_t-\bH^{-1}\bgamma_t
\|^2=   O_P(\frac{1}{N})  + O_P(\chi_N^{-1})(   \frac{ J^4(\log N)^2}{T^2} +\frac{J^2}{T^2}\chi_N^{-1}   +\frac{J\| \cov(\bgamma_s)\|}{T}+J^{1-2\eta}+\frac{J}{TN}).
$$

\section{Proof of Theorem \ref{t41}}
 The proof of the limiting distribution of $S$ under the null  is divided into two major steps.

 \textbf{step 1}: Asymptotic expansion: under $H_0$,
$$
S=\frac{1}{TN}\sum_{t=1}^T \bu_t'\bLambda\bH\widehat\bW\bH'\bLambda'\bu_t+o_P(T^{-1/2}).
$$

 \textbf{step 2}: The effect of estimating $\bSigma_u$ is first-order negligible:
 $$
\frac{1}{TN}\sum_{t=1}^T \bu_t'\bLambda\bH\widehat\bW\bH'\bLambda'\bu_t=\frac{1}{TN}\sum_{t=1}^T \bu_t'\bLambda (\frac{1}{N}\bLambda'\bSigma_u\bLambda)^{-1}\bLambda'\bu_t+ o_P(T^{-1/2}).
 $$
 The result then follows from the asymptotic normality of the first term on the right hand side. We shall prove this using Lindeberg's central limit theorem.   

 We achieve each step in the following subsections.

\subsection{Step 1 asymptotic expansion of $S$}

 \begin{prop}\label{pc.3} Under $H_0$,
$$
S=\frac{1}{TN}\sum_{t=1}^T \bu_t'\bLambda\bH\widehat\bW\bH'\bLambda'\bu_t+o_P(T^{-1/2})
$$
\end{prop}
 \proof
Since $\|\widehat\bW\|\leq\max_{i}\widehat\sigma_{ii}=O_P(1)$,  it  follows from (\ref{eb.2}) that it suffices to prove  under $H_0$,
$
\frac{N}{T}\sum_{t=1}^T\bD_{ti} '\widehat\bW\frac{1}{N}\bH'\bLambda'\bu_t=o_P(T^{-1/2}), $ and $ \frac{N}{T}\sum_{t=1}^T\|\bD_{ti}\|^2=o_P(T^{-1/2}), i=2,3,4.
$

By the proof of Propositions \ref{pb6}, \ref{pb3}, Lemmas \ref{lb3}, \ref{lb8} and that $\bD_{t3}=\bC_{t3}$,$\bD_{t4}=\bC_{t4}$,
\begin{eqnarray*}
\frac{N}{T}\sum_{t=1}^T\|\bD_{t4}\|^2  &=&O_P(\max_i\frac{N}{T}\sum_{t=1}^TR_{it}^2)\cr
&&=O_P(NJ^{1-2\eta}+ \frac{NJ^3\log N}{\alpha_T^{2(\zeta_1-1)}T}+\frac{NJ^3\log N\log J}{T^2})\cr
&=&o_P(\frac{1}{\sqrt{T}})\cr
\frac{N}{T}\sum_{t=1}^T\|\bD_{t3}\|^2 &=&O_P(\frac{1}{N}\|\widehat\bLambda'\bM_{\alpha}\|^2)\cr
&=&O_P(\frac{J}{T}+\frac{NJ\alpha_T^{-\zeta_2}}{T}+ J^{2-4\eta} +\alpha_T^{-2(\zeta_1-1)}\frac{J^3\log N}{TJ^{2\eta-1}})\cr
&=&o_P(\frac{1}{\sqrt{T}})
\end{eqnarray*}
The last equality holds so long as $N\sqrt{T}=o(J^{2\eta-1})$, $NJ^4\log N\log J=o(T^{3/2})$, $ \zeta_1>2$.

By Lemma \ref{lb7},
\begin{eqnarray*}
  \frac{N}{T}\sum_{t=1}^T\|\bD_{t2}\|^2&=&O_P(\frac{1}{N^2}\|\bM_{\alpha}'\widehat\bLambda	  \|^2+\max_i\frac{1}{T}\sum_{t=1}^T   R_{it}^2+ \frac{1}{NT} \sum_{s=1}^T\|\bu_s'	 \bM_{\alpha}\|^2 \cr
 &&+\frac{1}{NT^2} \sum_{s=1}^T         \sum_{t=1}^T| \bu_s'	\bR_t|^2)=o_P(\frac{1}{\sqrt{T}}).
\end{eqnarray*}
 The proof of $\frac{N}{T}\sum_{t=1}^T\bD_{ti} '\widehat\bW\frac{1}{N}\bH'\bLambda'\bu_t=o_P(T^{-1/2})$ is given in Lemmas  \ref{lc.2} and \ref{lc.3}.
It then leads to the desired result.

\subsection{Step 2 Completion of the proof}

We now aim to show $\widehat\bLambda'\widehat\bSigma_u\widehat\bLambda/N=\bH'\bLambda'\bSigma_u\bLambda\bH/N+o_P(T^{-1/2})$. Once this is done, it then follows from the facts that $\bH'\bLambda'\bSigma_u\bLambda\bH/N=O_P(1)$ and $(\bH'\bLambda'\bSigma_u\bLambda\bH/N)^{-1}=O_P(1)$,
$$(\widehat\bLambda'\widehat\bSigma_u\widehat\bLambda/N)^{-1}=(\bH'\bLambda'\bSigma_u\bLambda\bH/N)^{-1}+o_P(T^{-1/2}).$$
As a result, by Proposition \ref{pc.3},
\begin{eqnarray*}
S&=&\frac{1}{TN}\sum_{t=1}^T \bu_t'\bLambda\bH   (\bH'\bLambda'\bSigma_u\bLambda\bH/N)^{-1}   \bH'\bLambda'\bu_t+o_P(T^{-1/2})\cr
&=&\frac{1}{T}\sum_{t=1}^T \bu_t'\bLambda   (\bLambda'\bSigma_u\bLambda)^{-1}   \bLambda'\bu_t+o_P(T^{-1/2}).
\end{eqnarray*}
Hence
$$
\frac{TS-TK}{\sqrt{2TK}}=\frac{\sum_{t=1}^T \bu_t'\bLambda   (\bLambda'\bSigma_u\bLambda)^{-1}   \bLambda'\bu_t-TK}{\sqrt{2TK}}+o_P(1)\to^d\mathcal{N}(0,1).
$$

To finish the proof, we now show two claims: 

(1) 
 $$
\frac{ \sum_{t=1}^T \bu_t'\bLambda (\bLambda'\bSigma_u\bLambda)^{-1}\bLambda'\bu_t- TK}{\sqrt{2TK}}\to^d\mathcal{N}(0,1).
 $$
 
 (2)
  $\widehat\bLambda'\widehat\bSigma_u\widehat\bLambda/N=\bH'\bLambda'\bSigma_u\bLambda\bH/N+o_P(T^{-1/2})$.
  
  \textbf{Proof of (1)} We define $X_t=\bu_t'\bLambda (\bLambda'\bSigma_u\bLambda)^{-1}\bLambda'\bu_t$ and $s_T^2=\sum_{t=1}^T\var(X_t)$.  Then 
  $
  E(X_t)=\tr E( (\bLambda'\bSigma_u\bLambda)^{-1}\bLambda'\bu_t\bu_t'\bLambda)=K.
  $
Also    by Assumption 4.1, $s_T^2/T\to 2K$, hence we have  $E\frac{1}{T}\sum_{t=1}^T(X_t-K)^2<\infty$ for all large $N,T$.  For any $\epsilon>0$, by the dominated convergence theorem, for all large $N,T$, 
$$
\frac{1}{T}\sum_{t=1}^TE(X_t-K)^21\{|X_t-K|>\epsilon s_T\}\leq \frac{1}{T}\sum_{t=1}^TE(X_t-K)^21\{|X_t-K|>\epsilon \sqrt{KT}\}=o(1).
$$
  This then implies the Lindeberg condition, 
  $
  \frac{1}{s_T^2}\sum_{t=1}^TE(X_t-K)^21\{|X_t-K|>\epsilon s_T\}=o(1).
  $
  Hence by the Lindeberg central limit theorem, 
  $$
  \frac{\sum_tX_t-TK}{s_T}\to^d\mathcal{N}(0,1).
  $$
  The result then follows since $s_T^2/T\to 2K.$

  \textbf{Proof of (2)}
  By the triangular inequality,
\begin{eqnarray*}
 \|\frac{1}{N}\widehat\bLambda'\widehat\bSigma_u\widehat\bLambda-\frac{1}{N}\bH'\bLambda'\bSigma_u\bLambda\bH\|&\leq& \|\frac{1}{N}(\widehat\bLambda-\bLambda\bH)'(\widehat\bSigma_u-\bSigma_u)\widehat\bLambda\|\cr
&&+ \|\frac{1}{N}(\widehat\bLambda-\bLambda\bH)'\bSigma_u(\widehat\bLambda-\bLambda\bH)\|\cr
&&+\|\frac{1}{N}\bH'\bLambda'(\widehat\bSigma_u-\bSigma_u)(\widehat\bLambda-\bLambda\bH)\|\cr
&&+\|\frac{1}{N}\bH'\bLambda'(\widehat\bSigma_u-\bSigma_u)\bLambda\bH\|
\cr
&&+2\|\frac{1}{N}(\widehat\bLambda-\bLambda\bH)'\bSigma_u\bLambda\bH\|.
\end{eqnarray*}
Using the established bounds for $\|\widehat\bLambda-\bLambda\bH\|$   in Theorem 3.1, it is straightforward to verify $ \|\frac{1}{N}(\widehat\bLambda-\bLambda\bH)'\bSigma_u(\widehat\bLambda-\bLambda\bH)\|=o_P(T^{-1/2})$.  Other terms  require sharper bounds yet to be established. These are given in
Proposition \ref{lc.4}. It then follows  that  $\widehat\bLambda'\widehat\bSigma_u\widehat\bLambda/N=\bH'\bLambda'\bSigma_u\bLambda\bH/N+o_P(T^{-1/2})$. This completes the proof.

Q.E.D.

\newpage

  \bibliographystyle{ims}


\end{document}